\documentclass[a4paper,12pt]{article}
\usepackage{xcolor}
\usepackage{bbm}
\definecolor{nosaka}{rgb}{0.7, 0.3, 0.0}

\newcommand{\magica}{\includegraphics[height=4.000mm, trim = 0mm 0mm 0mm 0mm, clip]{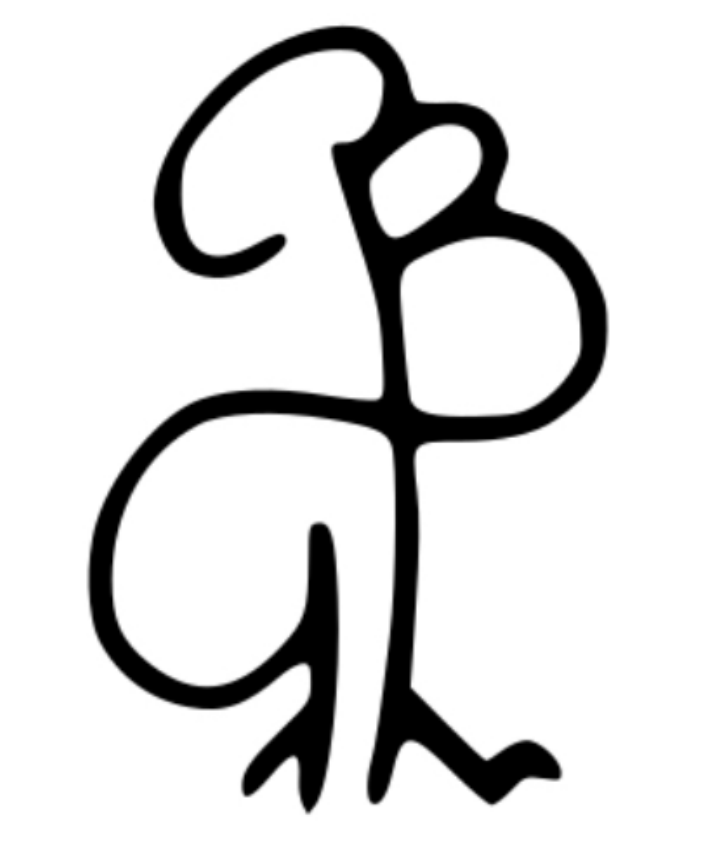}}
\newcommand{\magicb}{\includegraphics[height=3.750mm, trim = 0mm 0mm 0mm 0mm, clip]{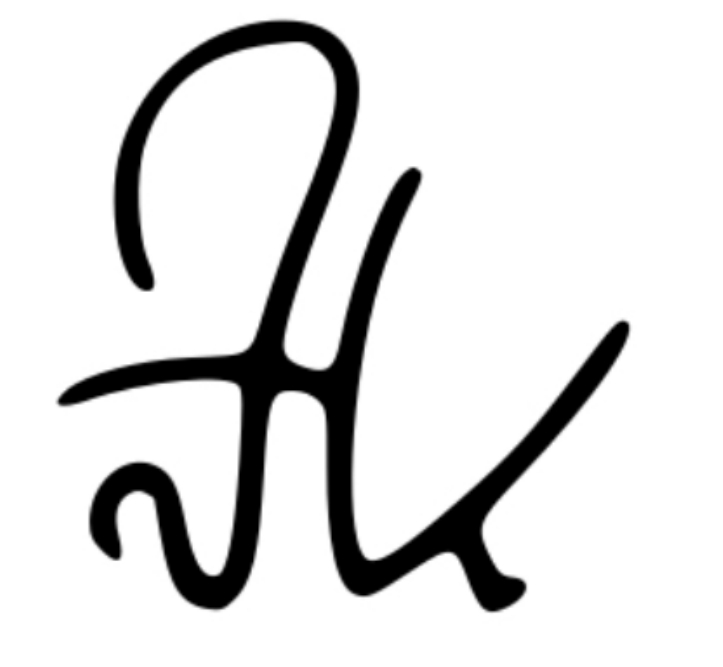}}
\usepackage{mathrsfs}
\usepackage{graphicx}
\usepackage{amsfonts,amsthm,amsmath,amssymb,graphicx,float,mathtools}
\numberwithin{equation}{section}
\usepackage{color}
\usepackage{ulem}
\usepackage{bbm} 
\usepackage{tensor} 
\usepackage{verbatim} 
\usepackage[
	  pagebackref=false,
	  colorlinks=true,
      linkcolor=blue,
      urlcolor=blue,
      filecolor=black,
      citecolor=red,
      pdfstartview=FitV,
      pdftitle={},
        pdfauthor={},
        pdfsubject={},
        pdfkeywords={},
        pdfpagemode=None,
        bookmarksopen=true
      ]{hyperref}

\usepackage{color,hyperref}
    \catcode`\_=11\relax
    
    \def\_email#1@#2\q_nil{%
      \href{mailto:#1@#2}{{\emailfont #1\emailampersat #2}}
    }
    \newcommand\emailfont{\sffamily}
    \newcommand\emailampersat{{\color{red}\small@}}
    \catcode`\_=8\relax    

\begin{document}

\newtheorem{definition}{Definition}[section]
\newcommand{\be}{\begin{equation}}
\newcommand{\ee}{\end{equation}}
\newcommand{\bea}{\begin{eqnarray}}
\newcommand{\eea}{\end{eqnarray}}
\newcommand{\LE}{\left[}
\newcommand{\R}{\right]}
\newcommand{\nn}{\nonumber}
\newcommand{\Tr}{\text{Tr}}
\newcommand{\N}{\mathcal{N}}
\newcommand{\G}{\Gamma}
\newcommand{\vf}{\varphi}
\newcommand{\LL}{\mathcal{L}}
\newcommand{\Op}{\mathcal{O}}
\newcommand{\HH}{\mathcal{H}}
\newcommand{\arctanh}{\text{arctanh}}
\newcommand{\up}{\uparrow}
\newcommand{\down}{\downarrow}
\newcommand{\ket}[1]{\left| #1 \right>}
\newcommand{\bra}[1]{\left< #1 \right|}
\newcommand{\ketbra}[1]{\left|#1\right>\left<#1\right|}
\newcommand{\rd}{\partial}
\newcommand{\de}{\partial}
\newcommand{\ba}{\begin{eqnarray}}
\newcommand{\ea}{\end{eqnarray}}
\newcommand{\db}{\bar{\partial}}
\newcommand{\we}{\wedge}
\newcommand{\ca}{\mathcal}
\newcommand{\lr}{\leftrightarrow}
\newcommand{\f}{\frac}
\newcommand{\s}{\sqrt}
\newcommand{\vp}{\varphi}
\newcommand{\hvp}{\hat{\varphi}}
\newcommand{\tvp}{\tilde{\varphi}}
\newcommand{\tp}{\tilde{\phi}}
\newcommand{\ti}{\tilde}
\newcommand{\ap}{\alpha}
\newcommand{\pr}{\propto}
\newcommand{\mb}{\mathbf}
\newcommand{\ddd}{\cdot\cdot\cdot}
\newcommand{\no}{\nonumber \\}
\newcommand{\la}{\langle}
\newcommand{\lb}{\rangle}
\newcommand{\ep}{\epsilon}
 \def\we{\wedge}
 \def\lr{\leftrightarrow}
 \def\f {\frac}
 \def\ti{\tilde}
 \def\ap{\alpha}
 \def\pr{\propto}
 \def\mb{\mathbf}
 \def\ddd{\cdot\cdot\cdot}
 \def\no{\nonumber \\}
 \def\la{\langle}
 \def\lb{\rangle}
 \def\ep{\epsilon}
\newcommand{\mcl}{\mathcal}
 \def\g{\gamma}
\def\Tr{\text{tr}}

\begin{titlepage}
\thispagestyle{empty}

\begin{flushright}
RIKEN-iTHEMS-Report-21
\end{flushright}
\bigskip

\begin{center}
  \noindent{{\Huge \textbf{Chaos {\it by} Magic
}}}\\
\vspace{1cm}
\renewcommand\thefootnote{\mbox{$\fnsymbol{footnote}$}}
Kanato Goto
${}^{\magica}$, Tomoki Nosaka
${}^{\magica,\magicb}$, and Masahiro Nozaki
${}^{\magica,\magicb}$\\

\vspace{1cm}

{\it \magica RIKEN Interdisciplinary Theoretical and Mathematical Sciences (iTHEMS),\\ Wako, Saitama 351-0198, Japan}

{\it \magicb Kavli Institute for Theoretical Sciences and CAS Center for Excellence in Topological Quantum Computation, University of Chinese Academy of Sciences, Beijing, 100190, China}
\vspace{1cm}
\vbox{\center\tt kanato.goto@riken.jp, nosaka@yukawa.kyoto-u.ac.jp, masahiro.nozaki@riken.jp}
\vskip 2em
\end{center}
\begin{abstract}
There is a property of a quantum state called magic.
It measures how difficult for a classical computer to simulate the state.
In this paper, we study magic of states in the integrable and chaotic regimes of the higher-spin generalization of the Ising model through two quantities called ``Mana'' and ``Robustness of Magic'' (RoM).
We find that in the chaotic regime, Mana increases monotonically in time in the early-time region, and at late times these quantities oscillate around some non-zero value that increases linearly with respect to the system size.
Our result also suggests that under chaotic dynamics, any state evolves to a state whose Mana almost saturates the optimal upper bound, i.e., the state becomes ``maximally magical.''
We find that RoM also shows similar behaviors.
On the other hand, in the integrable regime, Mana and RoM behave periodically in time in contrast to the chaotic case. In the anti-de Sitter/conformal field theory correspondence (AdS/CFT correspondence), classical spacetime emerges from the chaotic nature of the dual quantum system. Our result suggests that magic of quantum states is strongly involved behind the emergence of spacetime geometry.

\end{abstract}
\end{titlepage} 
\tableofcontents
\section{Introduction and Summary}
\subsection*{Introduction}

The question of how spacetime emerges from a more fundamental, microscopic concept of quantum gravity is one of the most intriguing questions in modern physics. In the context of holography, or the anti-de Sitter/conformal field theory correspondence (AdS/CFT correspondence) \cite{1999IJTP...38.1113M,1998,Witten:1998qj}, the properties of the classical spacetime emerge from the quantum nature of the dual CFTs. The connection between the quantum nature of the CFT and geometry of the dual spacetime has been examined by quantum information-theoretic notions such as ``quantum entanglement" \cite{2006PhRvL..96r1602R} and ``computational complexity" \cite{Susskind:2014moa}. In this paper, we would like to add a new quantum information-theoretic notion called ``magic" to this list, which can capture a classical feature of gravity as shown in the latter part of this paper\footnote{\cite{White:2020zoz} is a pioneering work on magic in a quantum system at the critical point. They studied magic in the $\mathbb{Z}_3$ Potts model and a tensor network model of AdS/CFT.}.

Quantum entanglement captures one aspect of the quantumness of a state, i.e., how much quantum correlation the state has. It is known that a CFT state with high energy density holographically corresponds to a black hole spacetime. Such a high energy density state, even if it is initially a simple state, quickly evolves into highly entangled, very complex states. In the dual geometry, this property of the state is captured by the growing black hole interior \cite{Hartman:2013qma}. This property of the classical spacetime can be measured by the holographic dual to entanglement entropy proposed in \cite{2006PhRvL..96r1602R} in most cases.

However, quantum entanglement alone does not fully account for the rich structure of a quantum state, nor does it fully characterize the properties of the dual classical spacetime. There is a situation where while the quantum entanglement reaches its equilibrium value relatively quickly, the black hole interior continues to expand afterward, and hence the quantum entanglement fails to capture the property of the classical spacetime \cite{Susskind:2014moa,Hartman:2013qma}.

It was proposed in \cite{Susskind:2014moa} that this continued growth in the black hole interior reflects the rowing ``computational complexity" of the quantum state \cite{2005quant.ph..2070N,2006Sci...311.1133N,2007quant.ph..1004D,Susskind:2014rva}, which entanglement entropy cannot capture. For simplicity, let us consider a discrete system that consists of $N$ qubits. A local unitary operator that interacts with only a few neighboring sites is called a gate.
A computational complexity is defined as the minimum number of gates required to build a target quantum state from a fixed reference state. In the chaotic system, one can expect that the  complexity of a state grows linearly in time since acting a gate per second on a state will give us a new state different from the original state. 

The holographic counterpart of the computational complexity for the CFT state dual to a AdS black hole was proposed in \cite{Brown:2015bva,Brown:2015lvg}. It is given by the gravitational action evaluated in the black hole interior (or more precisely, the so-called Wheeler-de Witt patch in the black hole spacetime)\footnote{Recent studies \cite{Goto:2018iay,Belin:2021bga} pointed out that there are a lot of ambiguities in this definition of the holographic complexity, but their details are out of scope of this paper.}. Since the size of the black hole interior measured by this action grows linearly in time, it correctly reproduces the typical behavior of complexity in chaotic systems.

``Magic," which we will study in this paper, measures complexity of quantum states that cannot be diagnosed by quantum entanglement or computational complexity.
To introduce magic, let us distinct all operations used in quantum computations into Clifford operations and the others. Clifford operations are a set of operations that can be efficiently simulated on a classical computer although some of them can generate quantum entanglement \cite{Gottesman:1997zz,Gottesman:1998hu}\footnote{Efficient here means that a quantum computer consisting of $n$-qubits can be computed in $n$ polynomial time on a classical computer.}. The states that they generate are called ``stabilizer states.'' This suggests that entanglement entropy, which measures only the amount of quantum entanglement, or computational complexity, which counts all gates equally, cannot distinguish whether a quantum state is indeed complex in the sense that it is difficult to perform classical calculations at high speed \footnote{For a related work, see also \cite{Fliss:2020yrd}.}. Such complexity is produced by the non-Clifford gates. We will explain the Clifford operations and stabilizer states in detail in sections \ref{stabilizer_state} and \ref{sec_stabilizerformalism}.

Magic is a property of a state that describes how difficult it is to simulate the state on a classical computer.
Roughly speaking, the difficulty to simulate a state $|\psi\rangle$ would be measured by how many copies of reference non-stabilizer states one need to prepare as an initial state to obtain $|\psi\rangle$ from the initial state through the Clifford operations.\footnote{
The minimum number of stabilizer states required to expand a quantum state is called the stabilizer rank $\chi$, and this $\chi$ is also one of the indicators for quantifying magic \cite{2016PhRvX...6b1043B,2016PhRvL.116y0501B}.
With this $\chi$, we can estimate the time it would take to simulate a quantum state on a classical computer.
}

In this paper, we evaluate magic through the quantities called ``Mana" and ``Robustness of Magic" (RoM). In particular, we are interested in the relation between magic and the chaotic property of a system.
The definitions of Mana and RoM are explained in sections \ref{D_o_Mana} and \ref{Def_o_RoM} respectively.
To this end, we study Mana and RoM in the higher-spin generalization of the Ising model, which has the chaotic and the integrable (non-chaotic) regime depending on the values of the parameters.

This model may not have a classical gravity dual since it only has small degrees of freedom and the dual gravity theory would become strongly coupled. However, for specific choices of the values of the parameters, as we explained, the dynamics of this system becomes chaotic and captures one of the characteristic properties of classical gravity. 

In this paper, we find that in the chaotic parameter regime, the states evolve to almost maximally magical states, suggesting that chaotic property is closely related to magic of a quantum state. From this result, we expect that magic is an important building block of classical spacetime, with the same footings as quantum entanglement and computational complexity.

\subsection*{Summary}
We studied the time evolution of pure-state Mana and RoM in the higher-spin generalization of the Ising model \cite{2020PhRvB.101q4313C}.
This model is integrable when the transverse magnetic field $h_x$ is turned off.
If the spin of each site is $J=1/2$, this model is integrable also when the longitudinal magnetic field $h_z$ vanishes while $h_x$ is non-zero.
When $(h_x,h_z)$ are not close to these points, on the other hand, the model is chaotic.
In the latter two cases, the chaotic property of the model is reflected well in the level statistics: when $J=1/2$ and $h_z=0$ the level statistics coincides with that of the set of independent random numbers, while when $J\ge 1$ and $h_z=0$ or when both of $h_x,h_z$ are non-zero (in particular around $(h_x,h_z)=(-1.05,0.5)$) the level statistics agrees with that of the random matrix theory with the Gaussian orthogonal ensemble (GOE).
We studied how the behavior of Mana and RoM depends on the dynamics of the system.

The main results obtained in the analysis of the time evolution of Mana and RoM in section \ref{TE_o_MM} are:

\begin{itemize}
\item[]
We found that the time evolution of Mana and RoM depends on the chaotic property of the dynamics.
At early times, both Mana and RoM increase monotonically, and at late times of the chaotic regime, Mana and RoM oscillate between the maximum values and the non-zero minimum values.
We found that these maxima and minima are respectively almost the same regardless of the choice of the initial state\footnote{
Here, we assumed that the initial state is not an energy eigenstate with low magic; in that case, magic of the state remains low at late times.
}.
On the other hand, in the integrable regime, both Mana and RoM behave periodically and depend on the initial states.
Hence
the typical minimum values of Mana and RoM at sufficiently late times are larger in the chaotic regime than in the integrable regime.
We also observed that in the chaotic regime, the late-time maximum value of Mana almost saturates the optimal bound that would be determined by the number of sites and the dimension of one-site Hilbert space.
As the number of sites increases, the typical minimum value of Mana at late times approximately coincides with the maximum value.
These results suggest that under the chaotic dynamics, any state evolves to a state which almost saturates the optimal bound of Mana.

\end{itemize}

The rest of the results in section \ref{TE_o_MM} is:

\begin{itemize}
\item[] 
We studied the time evolution of Mana and RoM when the open boundary condition and the periodic boundary condition were imposed on the system, and found that the late-time behavior of these quantities does not depend significantly on the boundary condition.
\end{itemize}

Holographic systems describing classical gravity have large degrees of freedom. It is difficult to make accurate predictions for such systems from those with small degrees of freedom we analyzed in this paper. However, once Mana is properly defined in the holographic systems, we expect that it encodes some geometrical information in anti-de Sitter space, as entanglement entropy and computational complexity do.
With this in mind, in section \ref{DaHM}, we added some comments on some possible behaviors of Mana in the holographic systems. We made some comments on the following three possibilities: 1.~Mana grows fast at early times and gradually slows down, as observed in the chaotic regime of the higher-spin generalized Ising model; 2.~Mana grows linearly in time until it saturates the upper bound derived from Jensen's inequality in section \ref{D_o_Mana} (and also its holographic version in section \ref{DaHM}); 3.~exponential of Mana, not Mana itself, grows linearly in time until it saturates the upper bound.
 In this section, we considered the time evolution of Mana for a quenched thermofield double state in a two-dimensional conformal field theory with gravity dual (so-called holographic CFT).
 It is known that this state describes a wormhole spacetime that continues to grow linearly with time via holography.
 
 Under some assumptions, we estimate the saturation time of Mana in cases 2 and 3.
 
 In case 1, the time dependence of Mana differs from the linear-growth of the wormhole captured by computational complexity (as well as entanglement entropy at early times). This suggests that the gravitational counterpart of Mana, even if it exists, is not a simple quantity to measure the size of the wormhole. 
 
 In case 2, we estimated that Mana would reach the upper bound in a polynomial time of the system size. In this case, the time dependence of Mana deviates from that of complexity at a relatively early time, suggesting that the state largely deviates from the ones that can be efficiently simulated on a classical computer after that time.
 
 In case 3, Mana would reach the upper bound in an exponential time of the system size. In this case, the time dependence of Mana does not deviate from that of complexity.

\subsection*{Organization}
So far, we have described the background of this study and the results obtained.
In section \ref{magic_in_qc}, we will explain the aspect of magic as the resource of quantum computation and the definition of two physical quantities, Mana and RoM, that measure magic.
In section \ref{Hsgim}, we describe the higher-spin generalized Ising model.
In section \ref{TE_o_MM}, we numerically study the time dependence of Mana and RoM in the chaotic and integrable regimes of the higher-spin generalized Ising model. We comment on the observations and some possible interpretations obtained from the numerical plots.
In section \ref{DaHM}, we comment on several possibilities of the behavior of Mana in the holographic systems. We estimate the time when Mana saturates its upper bound under some assumptions.
In section \ref{sec_discuss}, we discuss the results of this paper and some future directions.
\section{Magic in Quantum Computation \label{magic_in_qc}}
In this paper, we study how chaotic property emerges from quantum nature of a state through the notion of magic, which measures how difficult for a classical computer to simulate the state.
The purpose of this section is to define two quantities that measure magic of states, Mana and RoM, and to explain the resource aspect of magic.
We study the time evolution of these quantities in a higher-spin generalized Ising model in the chaotic and non-chaotic parameter regimes in section \ref{TE_o_MM}.
In section \ref{stabilizer_state}, we first introduce the Clifford operations, which are (parts of) the classical operations, and the stabilizer states generated by the Clifford operations.
Then, in sections \ref{D_o_Mana} and \ref{Def_o_RoM}, we introduce Mana and RoM.
Finally, in section \ref{sec_stabilizerformalism}, we will mention some aspects of magic in the framework of resource theory.

\subsection{Clifford group and stabilizer states}
\label{stabilizer_state}
To introduce the notion of magic, which measures how difficult for a classical computer to simulate the state, we first explain the Clifford group and the stabilizer states, the states which can be efficiently computed on a classical computer.

Let us consider a quantum system with $d$ orthonormal states (we assume $d$ to be a prime number) $|k\rangle$ ($k=0,1,\cdots,d-1$) and a pair of operators $z,x$ acting on this system
\begin{align}
z&=\sum_{k=0}^{d-1}\omega^k|k\rangle\langle k| \quad (\omega=e^{\frac{2\pi i}{d}}),\nonumber \\
x&=\sum_{k=0}^{d-1}|(k+1)\text{ mod }d\rangle\langle k|.
\label{zandx}
\end{align}
Note that $x,z$ satisfy the following properties
\begin{align}
x^d&=z^d=1,\quad x^az^b=\omega^{-ab}z^bx^a,\nonumber \\
\text{tr}z^ax^b&=\begin{cases}
d\quad &((a,b)=(0,0) \text{ mod }d)\\
0\quad &(\text{otherwise}),
\end{cases}
\end{align}
where $a$ and $b$ are integers.
We also define the generalized Pauli operators $t_{aa'}$ as
\begin{align}
t_{aa'}&=
\begin{cases}
i^{aa'}z^ax^{a'}\quad (d=2)\\
\omega^{-\bar{2}aa'}z^ax^{a'}\quad (d\ge 3)
\end{cases},
\end{align}
where $\bar{2}$ is an integer such that $2\times \bar{2}\equiv 1$ mod $d$.
For general $d$, $t_{aa'}$ satisfy the following relations
\begin{align}
t_{aa'}^\dagger&=t_{-a,-a'},\quad
t_{aa'}t_{bb'}=\omega^{\bar{2}(ab'-ba')}t_{a+b,a'+b'}=\omega^{ab'-ba'}t_{bb'}t_{aa'},\label{noncommutativityoft} \\
\text{tr}t_{aa'}&=\begin{cases}
d\quad &((a,a')=(0,0) \text{ mod }d)\\
0\quad &(\text{otherwise})
\end{cases}.
\end{align}
In particular, $t_{aa'}$ are orthonormalized in the following sense:
\begin{align}
\text{tr}\left(t_{aa'}t_{bb'}^\dagger\right)=\begin{cases}
d\quad &((a,a')=(b,b') \text{ mod }d)\\
0\quad &(\text{otherwise})
\end{cases}.
\end{align}
We consider a system that consists of $L$ sites of this generalized qubit (or ``qudit'') whose Hilbert space is spanned by $\{|k_1\rangle \otimes |k_2\rangle\otimes \cdots \otimes |k_L\rangle\}$, and consider the generalized Pauli operators (generalized Pauli strings) acting on these states labeled by\\ $\overrightarrow{a}=((a_1,a'_1),(a_2,a'_2),\cdots,(a_L,a'_L))$:
\begin{align}
T_{\overrightarrow{a}}=
t_{a_1a_1'}\otimes
t_{a_2a_2'}\otimes
\cdots
\otimes t_{a_La_L'}.
\label{211112generalizedPaulistrings}
\end{align}

The Clifford group $C_d$ is a discrete set of unitary matrices such that each element of $C_d$ transforms the set of all Pauli strings $\{T_{\overrightarrow a}\}$ to itself up to some overall phases:
\begin{align}
U\in C_d\Leftrightarrow
U T_{\overrightarrow a} U^\dagger = e^{i\phi_U({\overrightarrow a})} T_{\sigma(\overrightarrow a)}
\text{ for all }\overrightarrow a,
\end{align}
where $e^{i\phi_U({\overrightarrow a})}$ are some phases and $\sigma_U$ is a permutation on $d^L$ choices of ${\overrightarrow a}$.
The symbol $\phi$ is a real number.
Explicitly, the Clifford group $C_d$ for $d=2$ is generated by the following elements
$R_a,P_a,\text{SUM}_{a,b}$ which are respectively called Hadamard gate, phase gate and controlled-NOT gate\footnote{
Here, we follow the notation of \cite{Gottesman:1997qd}.
} \cite{Gottesman:1997qd}:
\begin{itemize}
\item $R_a$,$P_a$ ($a=1,2,\cdots,L$):
\begin{align}
R_a&=1\otimes \cdots 1\otimes \mathop{R}_a\otimes 1\otimes \cdots 1,\nonumber \\
P_a&=1\otimes \cdots 1\otimes \mathop{P}_a\otimes 1\otimes \cdots 1,
\end{align}
with
\begin{align}
R=
\frac{1}{\sqrt{2}}
\begin{pmatrix}
1&1\\
1&-1
\end{pmatrix},\quad
P=\begin{pmatrix}
1&0\\
0&i
\end{pmatrix},
\end{align}
\item $\text{CNOT}_{a,b}$ ($a,b=1,2,\cdots,L$):
\begin{align}
\text{CNOT}_{a,b}:(\cdots \otimes \mathop{|k_a\rangle}_a\otimes\cdots\mathop{|k_b\rangle}_b\cdots)\rightarrow
(\cdots \otimes \mathop{|k_a\rangle}_a\otimes\cdots\mathop{|(k_a+k_b)\text{ mod }2\rangle}_b\cdots),
\end{align}
where $|k\rangle$ is the eigenstate of $z$ with eigenvalue $z=\omega^k$, as introduced above \eqref{zandx}. 
\end{itemize}
On the other hand, the Clifford group $C_d$ for $d>3$ is generated by the following elements $R_a,P_a,S_a,\text{SUM}_{a,b}$ \cite{Gottesman:1998se}:
\begin{itemize}
\item $R_a,P_a,S_a$ ($a=1,2,\cdots,L$):
\begin{align}
R_a&=1\otimes \cdots 1\otimes \mathop{R}_a\otimes 1\otimes \cdots 1,\nonumber \\
P_a&=1\otimes \cdots 1\otimes \mathop{P}_a\otimes 1\otimes \cdots 1,\nonumber \\
S_a&=1\otimes \cdots 1\otimes \mathop{S}_a\otimes 1\otimes \cdots 1
\end{align}
with
\begin{align}
R&=\frac{1}{\sqrt{d}}(\omega^{(k-1)(\ell-1)})_{0\le k,\ell\le d-1}
=\frac{1}{\sqrt{d}}\begin{pmatrix}
1&1       &1       &\cdots&1\\
1&\omega  &\omega^2&\cdots&\omega^{d-1}\\
1&\omega^2&\omega^4&\cdots&\omega^{2(d-1)}\\
 &        &\vdots  &      &
\end{pmatrix},\nonumber \\
P&=(\omega^{\frac{k(k-1)}{2}}\delta_{k\ell})_{0\le k,\ell\le d-1}
=\begin{pmatrix}
1&0     &\cdots  &        &      &                             0\\
0&\omega&0       &\cdots  &      &                             0\\
0&0     &\omega^3&0       &\cdots&                             0\\
 &      &        &\ddots&                             &\\
0&\cdots&        &0     &\omega^{\frac{(d-2)(d-3)}{2}}&0\\
0&\cdots&        &      &0                            &\omega^{\frac{(d-1)(d-2)}{2}}
\end{pmatrix},\nonumber \\
S&=(\delta_{k-a\ell,d (\text{mod }d)})_{1\le k,\ell\le d-1},
\label{210820RPS}
\end{align}
where $a$ in the right-hand side of $S$ is an integer such that $\{a,a^2,a^3,\cdots\} (\text{mod }d)\supset (\mathbb{Z}_d\backslash \{0\})$.
One can choose any $a$ satisfying this condition.
For example, for $d=3$, the choice $a=2$ works since all the element of $\mathbb{Z}_d\backslash \{0\}=\{1,2\}$ can be realized as $a^2=4\equiv 1$, $a=2$.
\item SUM gate $\text{SUM}_{a,b}$ ($a,b=1,2,\cdots,L$):
\begin{align}
\text{SUM}_{a,b}:(\cdots \otimes \mathop{|k_a\rangle}_a\otimes\cdots\mathop{|k_b\rangle}_b\cdots)\rightarrow
(\cdots \otimes \mathop{|k_a\rangle}_a\otimes\cdots\mathop{|(k_a+k_b)\text{ mod }d\rangle}_b\cdots).
\label{210820SUM}
\end{align}
\end{itemize}

Choose one of the eigenstates of any single generalized Pauli operator as a base state.
Here, let us choose $|{\overrightarrow 0}\rangle=|0\rangle\otimes |0\rangle\otimes \cdots \otimes |0\rangle$ as the base state.
The set of all stabilizer states $\{|S\rangle\}$ is defined as the states generated by acting the elements of the Clifford group $C_d$ on the base state:
\begin{align}
\{|S\rangle\}=\{U|{\overrightarrow 0}\rangle\,|\,U\in C_d\}.
\label{211112stabilizerpurestates}
\end{align}
For general $d$ and $L$, the number of all stabilizer states is given as \cite{2006JMP....47l2107G}
\begin{align}
|\{|S\rangle\}|=d^L\prod_{n=1}^L(d^n+1).
\label{211221numberofstabilizerpurestates}
\end{align}
We also define the convex hull of the stabilizer pure states $\{|s\rangle\langle s|\}_{|s\rangle\in\{|S\rangle\}}$, which we call STAB:
\begin{align}
\text{STAB}=\Bigl\{\sum_ic_i|s_i\rangle \bra{s_i}\,\Bigl|\,|s_i\rangle\in\{|S\rangle\},c_i\ge 0,\sum_ic_i=1\Bigr\}.
\label{211112STAB}
\end{align}

\subsection{Mana}
\label{D_o_Mana}
In this section, we introduce one of the measures for magic of a quantum state called Mana.
Since STAB is a convex hull spanned by the stabilizer pure states, it would be natural to expect that for a given state expanded in the basis of stabilizer states, the negative coefficients in the expansion can be used to quantify the discrepancy of the state from STAB.
Indeed, both Mana and Robustness of Magic are related to the negativity of the state in this sense.

To define Mana, we first define the phase space point operator $A_{\overrightarrow{a}}$ by using the generalized Pauli strings \eqref{211112generalizedPaulistrings} as
\begin{align}
A_{\overrightarrow{a}}=d^{-L}T_{\overrightarrow{a}}\sum_{\overrightarrow{b}}T_{\overrightarrow{b}}T_{\overrightarrow{a}}^\dagger,
\end{align}
which satisfies\footnote{
Note that by using \eqref{noncommutativityoft}, we can also write $A_{\overrightarrow{a}}$ as
\begin{align}
A_{\overrightarrow{a}}=d^{-L}\bigotimes_{i=1}^L\biggl(\sum_{b,b'}\omega^{a_ib'-a'_ib}t_{bb'}\biggr).
\end{align}
This expression is useful to show \eqref{hermiteorthonormalA}.
}
\begin{align}
A_{\overrightarrow{a}}^\dagger=A_{\overrightarrow{a}},\quad
\text{Tr}A_{\overrightarrow{a}}A_{\overrightarrow{b}}=\begin{cases}
d^L&\quad (\overrightarrow{a}=\overrightarrow{b}\text{ mod }d)\\
0&\quad (\text{otherwise})
\end{cases}.
\label{hermiteorthonormalA}
\end{align}
With these phase space point operators $A_{\overrightarrow{a}}$, we define discrete Wigner functions $W_\rho(\overrightarrow{a})$ of a given density state $\rho$ as
\begin{align}
W_\rho(\overrightarrow{a})=\frac{1}{d^L}\text{Tr}\rho A_{\overrightarrow{a}}.
\end{align}
Since a set of the phase space point operators $\{A_{\overrightarrow a}\}$ forms a complete orthonormal basis of $d^L\times d^L$ Hermitian matrices, this implies $\rho=\sum_{\overrightarrow a}W_\rho({\overrightarrow a})A_{\overrightarrow a}$.
 If we impose the normalization condition $\text{tr}\rho=1$, $\{W_\rho({\overrightarrow a})\}$ satisfies $\sum_{\overrightarrow a}W_\rho({\overrightarrow a})=1$.
Then, we define Mana $M(\rho)$ of a state $\rho$ as the negativity of $\{W_\rho(\overrightarrow{a})\}$:
\begin{align}
M(\rho)=\log\sum_{\overrightarrow{a}}|W_\rho(\overrightarrow{a})|=\log \Biggl[1+2\sum_{\substack{\overrightarrow{a}\\ (\text{s.t.}\,W_\rho(\overrightarrow{a})<0)}}|W_\rho(\overrightarrow{a})|\Biggr].
\label{Manaforqudit}
\end{align}
For $\rho$ being a pure state, it is known that $M(\rho)=0$ if and only if $\rho$ is a stabilizer pure state (discrete Hudson's theorem) \cite{2006JMP....47l2107G}.
Since STAB is the convex hull of the stabilizer pure states, this also implies that $M(\rho)=0$ for any $\rho\in\text{STAB}$.
However, when $\rho$ is a mixed state, $M(\rho)=0$ does not necessarily imply $\rho\in\text{STAB}$.
In this sense, Mana is not a faithful quantification of the non-stabilizerness of the states.
Also, Mana is defined only when the number of states $d$ on a single site is odd.
Nevertheless, Mana is a useful measure to quantify magic of the state since the definition of Mana does not involve any optimization over the large degrees of freedom, unlike Robustness of Magic \eqref{211112RoM} or the relative entropy of magic \cite{2014NJPh...16a3009V} (which we do not consider in this paper).

Note that Mana $M(\rho)$ obeys the following inequality
\begin{align}
M(\rho)\le \frac{1}{2}(L\log d-S_2),
\label{Jensen}
\end{align}
where $S_2$ is the second R\'{e}nyi entropy $S_2=-\log\text{tr}\rho^2$ which can also be written in $W_\rho({\overrightarrow a})$ as
\begin{align}
e^{-S_2}=d^L\sum_{\overrightarrow{a}}W({\overrightarrow a})^2.
\end{align}
The inequality \eqref{Jensen} can be shown by using the concavity of the function $f(x)=\sqrt{x}$ (Jensen's inequality). For more details of the derivation of the inequality, see \cite{2020arXiv201113937W}.

Let us add some comments on the upper bound for the case where $\rho$ is a pure state, $\rho=|\psi\rangle\langle\psi|$, which is of our main interest in the subsequent sections.
In this case, Jensen's inequality reduces to
\begin{align}
M(|\psi\rangle\langle\psi|)\le \frac{L}{2}\log d.
\label{Jensenpure}
\end{align}
Note that the inequality \eqref{Jensenpure} (also \eqref{Jensen}) does not mean that there actually exists either a state which saturates the bound, or an infinite series of states whose Mana converges to the saturation of the bound.
That is, the right-hand side of \eqref{Jensenpure} may not be the optimal upper bound, which we shall call $M_0(d,L)$, on Mana of the pure states in the qudit system with $L$ sites.
Indeed for $d=3$ and $L=1$, $M_0(d,L)$ is known to be \cite{2014NJPh...16a3009V}\footnote{
The optimal upper bound is also known for $d=5$ and $L=1$ as $M_0(5,1)=\text{Arcsinh}(3+\sqrt{5})-\log 5$ \cite{2020PhRvA.102d2409J}, which is smaller than $(1/2) \log 5$, the right-hand side of \eqref{Jensenpure}.
However, we will not use this result in this paper.
}
\begin{align}
M_0(3,1)=\log\Bigl(\frac{5}{3}\Bigr),
\label{d3L1optimal}
\end{align}
which is smaller than $(1/2) \log 3$.
To our knowledge, the optimal bound for $L\ge 2$ is still not known.
Nevertheless, for $d=3$, since there exists a state $|\psi\rangle=|\mathbb{S}\rangle^{\otimes L}$ where $|\mathbb{S}\rangle=|\mathbb{S}\rangle=(|1\rangle-|2\rangle)/\sqrt{2}$ is the state in the single qudit system with $M(|\mathbb{S}\rangle\langle\mathbb{S}|)=M_0(3,1)$ (see \cite{2014NJPh...16a3009V}), whose Mana is $M(|\psi\rangle\langle\psi|)=LM(|\mathbb{S}\rangle\langle\mathbb{S}|)=LM_0(3,1)$, the optimal upper bound $M_0(3,L)$ must not be less than $LM_0(3,1)$.
Together with \eqref{Jensenpure}, we conclude that $M_0(3,L)$ is in the following window:
\begin{align}
L\log\Bigl(\frac{5}{3}\Bigr)\le M_0(3,L)\le\frac{L}{2}\log 3.
\label{window}
\end{align}
In section \ref{time_evo_mana}, we will use this fact to interpret the numerical results in the chaotic regime of the higher-spin generalized Ising model.

\subsection{Robustness of Magic}
\label{Def_o_RoM}
We have defined Mana, one of the  measures for magic of a quantum state.
Here, we will introduce another measure called Robustness of Magic. 
Robustness of Magic, $\text{RoM}(\rho)$, of a given state $\rho$ is defined as \cite{2017PhRvL.118i0501H,2018PhRvA..97f2332A,2018arXiv180710296H,2020NJPh...22h3077S}
\begin{align}
\text{RoM}(\rho)=\text{inf}\Bigl\{
\sum_{\{|S\rangle\}\, (V_{|S\rangle}<0)}|V_{|S\rangle}|\quad \big{|}\quad
V_{|S\rangle}\in\mathbb{R},\quad 
\sum_{\{|S\rangle\}}B_{{\overrightarrow a},|S\rangle}V_{|S\rangle}=F_{{\overrightarrow a}}(\rho)
\Bigr\},
\label{211112RoM}
\end{align}
where
\begin{align}
B_{{\overrightarrow a},|S\rangle}=\text{Tr}(T_{{\overrightarrow a}}|S\rangle\langle S|),\quad
F_{{\overrightarrow a}}=\text{Tr}(T_{{\overrightarrow a}}\rho),
\end{align}
 with the generalized Pauli strings $T_{{\overrightarrow a}}$ \eqref{211112generalizedPaulistrings} and the stabilizer pure states $|S\rangle$ \eqref{211112stabilizerpurestates}.
Note that since $\{T_{{\overrightarrow a}}\}$ is a complete set of $d^L\times d^L$ matrices, the constraint $\sum_{{\overrightarrow b}}B_{{\overrightarrow a},|S\rangle}V_{|S\rangle}=F_{{\overrightarrow a}}(\rho)$ is equivalent to the following
\begin{align}
\rho=\sum_{\{|S\rangle\}}V_{|S\rangle}|S\rangle\langle S|.
\end{align}
That is, $\text{RoM}(\rho)$ \eqref{211112RoM} directly measures the amount of the negative coefficients when $\rho$ is expanded in $\{|s\rangle\langle s|\}_{|s\rangle\in\{|S\rangle\}}$ in the most optimal way.

\subsection{Stabilizer formalism and magic monotone}
\label{sec_stabilizerformalism}

We would also like to briefly comment on the stabilizer formalism and the resource theory of magic, which allows us to formulate the ``non-stabilizerness'' systematically.

Let us start with a brief explanation of the resource theory.
The resource theory is an idea to classify the quantum states by using the following three notions: (i) a set of operations ${\cal C}$, which in general are non-invertible; (ii) free states ${\cal S}$, which can be created from any single state in ${\cal S}$ by acting full ${\cal C}$ on it; (iii) monotone, which is some quantity defined for any states so that it does not increase under ${\cal C}$ and that it vanishes for the elements in ${\cal S}$.
The states which cannot be generated by acting ${\cal C}$ on any free states are called resource states.
A monotone is positive if (provided that the monotone is faithful) and only if the state in concern is a resource state.
From the definition of the monotones, it also follows that it is impossible to obtain a resource state by acting ${\cal C}$ on another resource state with a smaller monotone.
In this sense, a monotone of a state $\rho$ quantifies the diversity of the states that can be generated from $\rho$ by acting ${\cal C}$.

As a concrete example, let us explain the entanglement and the related notions, which will be more familiar to the reader, in the language of the resource theory \cite{RevModPhys.81.865,2011arXiv1111.3882B,2013IJMPB..2745019H,2019RvMP...91b5001C}.
In the resource theory of entanglement, the role of ${\cal C}$ is played by the set of operations which consists of local operations and classical communication (LOCC) as operations to solve quantum entanglement of states.
The free states ${\cal S}$ corresponds to the set of separable density matrices.
Conversely, the resource states are the entangled states.
A physical quantity with the properties of the motonone in the resource theory of entanglement is called an entanglement monotone \cite{PhysRevA.53.2046,2000JMOp...47..355V}.
The entanglement entropy is a typical example of the entanglement monotones for the pure states.

In the context of magic, the set of operations ${\cal C}$ corresponds to the stabilizer protocols, and the free states ${\cal S}$ corresponds to STAB \eqref{211112STAB}.
The stabilizer protocol is the set of operations that consists of the Clifford group and some additional operations (composition of any stabilizer state; projection measurement on a single site into any of the computational basis $|k\rangle$; a partial trace of a single site) \cite{2014NJPh...16a3009V}.
The image of the stabilizer protocol acting on a set of states is in general smaller than the initial set, while the image of STAB is itself.
These properties suggest that we can consider the stabilizer protocol and STAB as the elements of the resource theory \cite{2013IJMPB..2745019H,2014NJPh...16a3009V}.
A monotone of a given resource state indicates how many states can be generated by the stabilizer operations on this state. 
It also quantifies how many copies of the given resource state $|\psi\rangle$ are required to generate a fixed target resource state $|\phi\rangle$ (with a large monotone) by ${\cal C}$. This is a kind of the efficiency of the universal quantum computation realized by the stabilizer operations together with that resource state $|\psi\rangle$, called stabilizer formalism \cite{2014NJPh...16a3009V}.
In \cite{2014NJPh...16a3009V}, the monotone and the resource states of the stabilizer operations are particularly called the ``magic monotone'' and the ``magic states''.
Mana and RoM are examples of the magic monotone.

\section{Higher-Spin Generalized Ising Model\label{Hsgim}}

As an example of the qudit system with adjustable quantum chaoticity, we consider the higher-spin generalized Ising model with the open/periodic boundary condition \cite{2020PhRvB.101q4313C}:
\begin{align}
H=
\begin{cases}
\frac{2}{\sqrt{3}}\sqrt{J(J+1)}\Bigl[-\sum_{n=1}^{L-1}G_z^{(n)}G_z^{(n+1)}-\sum_{n=1}^L(h_xG_x^{(n)}+h_zG_z^{(n)})\Bigr]\quad \text{(open)}\vspace{0.2cm}\\
\frac{2}{\sqrt{3}}\sqrt{J(J+1)}\Bigl[-\sum_{n=1}^{L-1}G_z^{(n)}G_z^{(n+1)}-G_z^{(L)}G_z^{(1)}-\sum_{n=1}^L(h_xG_x^{(n)}+h_zG_z^{(n)})\Bigr]\quad \text{(periodic)}
\end{cases},
\label{HhigherspinIsing}
\end{align}
where $G_i^{(n)}$ are given as
\begin{align}
G_i^{(n)}=\mathop{1}_1\otimes 1\otimes \cdots\otimes 1\mathop{G_i}_n\otimes 1\otimes 1\otimes \cdots\otimes \mathop{1}_L,
\end{align}
with $G_i$ being the $2J+1$ dimensional representation of $SU(2)$ generators. 
If we choose the eigenstates of $G_z$, $\{|j=J\rangle,|j=J-1\rangle,\cdots,|j=-J\rangle\}$, as the basis of the single-site Hilbert space, then $G_i$ take the following form
\begin{align}
&G_z=\begin{pmatrix}
J\\
&J-1\\
&&\ddots\\
&&&-J
\end{pmatrix},\quad
G_x=\frac{1}{\sqrt{2}}\begin{pmatrix}
0  &a_0   &0     &\cdots&      &      &0\\
a_0&0     &a_1   &0     &\cdots&      &0\\
0  &a_1   &0     &a_2   &0     &\cdots&0\\
   &\ddots&\ddots&\ddots&\ddots&
\end{pmatrix},\nonumber \\
&a_i=\sqrt{\frac{(2J-i)(i+1)}{2}}.
\label{SzandSx}
\end{align}

\subsection{Chaotic property}
The model \eqref{HhigherspinIsing} is trivially integrable when $h_x=0$, while when $h_x\neq 0$ the chaotic property of this model varies depending on the values of $h_x,h_z$ (and also on $J$).
There are several different ways to characterize the chaoticity of quantum many-body systems.
Here, let us adopt, as a diagnosis of quantum chaos, the nearest-neighbor spacing distribution (NNSD) of the energy spectrum $\{E_n\}_{n=1}^{\text{dim}{\cal H}}$ \cite{1984LNP...209....1B,Guhr:1997ve} which is the value distribution of $\{E_n-E_{n-1}\}$ normalized by the average value of $E_n-E_{n-1}$ over $n$.
It has been observed that
the NNSD of the energy spectrum obtained by quantizing a classical integrable system obeys the Poisson distribution \cite{1977RSPSA.356..375B}, while the energy spectrum obtained by quantizing a chaotic system shows a similar NNSD to that of the random matrix theory (RMT) with the ensemble determined by the time-reversal property of the system \cite{PhysRevLett.52.1}.
This characterization has also been found to be consistent with other characterizations using level statistics such as the onset of the RMT-like linear growth (ramp) of the spectral form factor and also with the characterization by the growth exponent of the connected part of the out-of-time-ordered correlator (quantum chaos exponent) \cite{1969JETP...28.1200L}, provided that all of these quantities in comparison are well defined \cite{Nosaka:2018iat,Kudler-Flam:2019kxq}.

To characterize the chaoticity of a system correctly by using NNSD, we have to follow the following prescriptions.
First, if the system enjoys some discrete symmetry, we have to split the full spectrum into the irreducible sectors protected by the symmetry and define the level spacing distribution by using only the energy levels within each sector.
Second, if the averaged eigenvalue density $\bar{\rho}(E)$ of each sector is not uniform, we have to perform a redefinition of eigenvalues ${\widetilde E}_n=\int_{E_0}^{E_n} \bar{\rho}(E')dE'$ called unfolding \cite{Guhr:1997ve} and define the level spacing as ${\widetilde E}_n-{\widetilde E}_{n-1}$ instead of $E_n-E_{n-1}$.

In the higher-spin generalized Ising model with the open boundary condition \eqref{HhigherspinIsing}, there is the reflection symmetry ${\cal O}_P$ which exchanges $i$-th site with $(L-i)$-th site.
Hence we have to split the energy spectrum into those of reflection-even eigenstates ${\cal O}_P|n\rangle=|n\rangle$ and those of reflection-odd eigenstates ${\cal O}_P|n\rangle=-|n\rangle$.
This can be done, for example, by deforming the Hamiltonian as
\begin{align}
H\rightarrow H+\Lambda {\cal O}_P
\end{align}
with $\Lambda$ any number sufficiently larger than the original bandwidth, and collecting the energy levels around $E\sim \pm \Lambda$.
\begin{figure}
\begin{center}
\includegraphics[width=8cm]{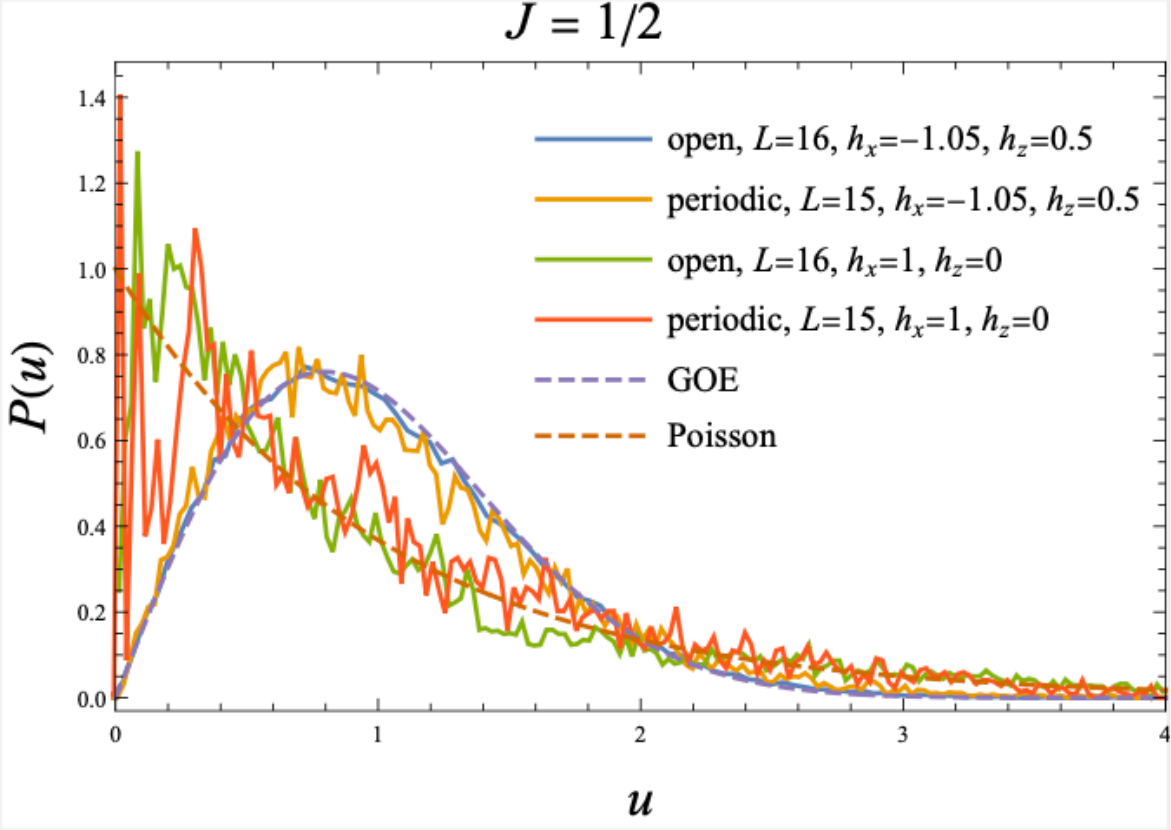}
\includegraphics[width=8cm]{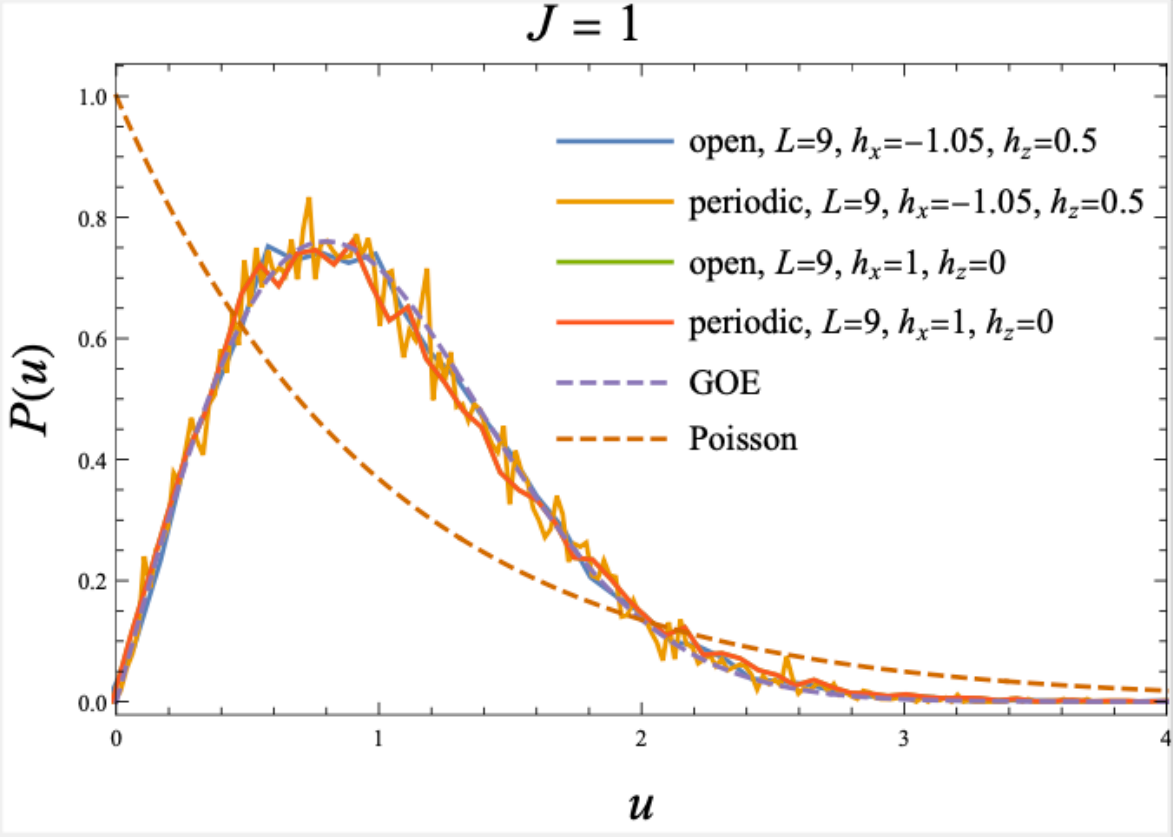}
\caption{The NNSD of the higher-spin generalized Ising model with the open/periodic boundary condition \eqref{HhigherspinIsing}.
Here, $u$ in the horizontal axis is the values of ${\widetilde E}_n-{\widetilde E}_{n-1}$ normalized by the average over the subsector in concern.
Each histogram is drawn by first computing the level spacings of the unfolded spectrum, normalizing in each sector separately, and then combining the results of all the subsectors.
We chose the background magnetic fields as $h_x=-1.05, h_z=0.5$ and $h_x=1, h_z=0$.
}
\label{NNSDopen}
\end{center}
\end{figure}
For the periodic case, there is also the translation symmetry ${\cal O}_T$ which changes $i$-th site into $((i+1)\text{ mod }L)$-th site.
Since ${\cal O}_T$ does not commute with ${\cal O}_P$, a sector labeled by the eigenvalue of ${\cal O}_T$ does not always split to the ${\cal O}_P$-protected sectors.
However, the ${\cal O}_T$-protected sector with ${\cal O}_T=0$ transforms to itself under the reflection, hence this sector can be further decomposed into the ${\cal O}_P$-protected subsectors.\footnote{
When $L$ is even, this also applies to the ${\cal O}_T=-1$ sector.
}
Lastly, when we take the parameter $h_z$ to be zero, the Hamiltonian (both the open/periodic boundary condition) is invariant under an additional transformation ${\cal O}_Y: (G_x^{(n)},G_z^{(n)})\rightarrow (G_x^{(n)},-G_z^{(n)})$.
The explicit expressions for the unitary matrix ${\cal O}_Y$ which realizes this symmetry transformation for $J=1/2,1$ are the followings:
\begin{align}
{\cal O}_Y(d=2)=\begin{pmatrix}
0&-i\\
i&0
\end{pmatrix}^{\otimes L},\quad
{\cal O}_Y(d=3)=\begin{pmatrix}
0&0&1\\
0&1&0\\
1&0&0
\end{pmatrix}^{\otimes L}.
\end{align}
This unitary matrix ${\cal O}_Y$ has the eigenvalues $\pm 1$ and commutes with the geometrical symmetry transformations ${\cal O}_P$ and ${\cal O}_T$.

By unfolding each sector, we obtained NNSD as displayed in figure \ref{NNSDopen}.
The results indicate that the higher-spin generalized Ising model is chaotic for $(J,h_x,h_z)=(1/2,-1.05,0.5)$ and $(1,1,0)$ \cite{2011PhRvL.106e0405B,2020PhRvB.101q4313C}, while the model is integrable for $(J,h_x,h_z)=(1/2,1,0)$ \cite{2011PhRvL.106e0405B}.

\section{Time Dependence of Mana and RoM\label{TE_o_MM}}

The main aim of this paper is to elucidate how the chaotic property emerges from the quantum nature of a system through the notion of magic. For this purpose, we study the time evolution of Mana and RoM introduced in the previous section for the chaotic and non-chaotic regimes of the  higher-spin generalized Ising model.

In this paper, we identify the eigenstates of $G_i$, $|J\rangle,|j=J-1\rangle,\cdots,|j=-J\rangle$ with the computational basis $|k\rangle$ in \eqref{zandx} as $|k\rangle=|j=J+1-k\rangle$, and choose the stabilizer states $\left| S\right\rangle$ as the initial states \eqref{211112STAB} and evolve it under the Hamiltonian of the system: $e^{-iHt}|S\rangle$.
For simplicity\footnote{
We chose a state which respects the reflection symmetry which exists for any choice of $(J,h_x,h_z)$ to make the computation easier.
At the same time, we avoided a state which reflects any of the enhanced global symmetry $T,U$ which appears only when the system is periodic or when $h_x=0$ so that the comparison of the results for the different chaotic models (open, open with $d=3, h_x=0$, closed, closed with $d=3, h_x=0$) would be more reasonable.
},
we choose $|x=\omega\rangle\otimes |x=1\rangle^{\otimes (L-2)}\otimes |x=\omega\rangle$, where $|x=\omega^n\rangle$ is the eigenstate of $x$ \eqref{zandx} with eigenvalue $\omega^n$, as the initial stabilizer pure state $\ket{S}$. We will abbreviate this state as $|x1x0\cdots x0x1\rangle$.
In sections \ref{time_evo_mana} and \ref{time_evo_RoM}, we display the results of the analysis.
Lastly, in section \ref{Observe}, we list the observations and their interpretations.

\subsection{Mana \label{time_evo_mana}}
\begin{figure}
\begin{center}
\includegraphics[width=16cm]{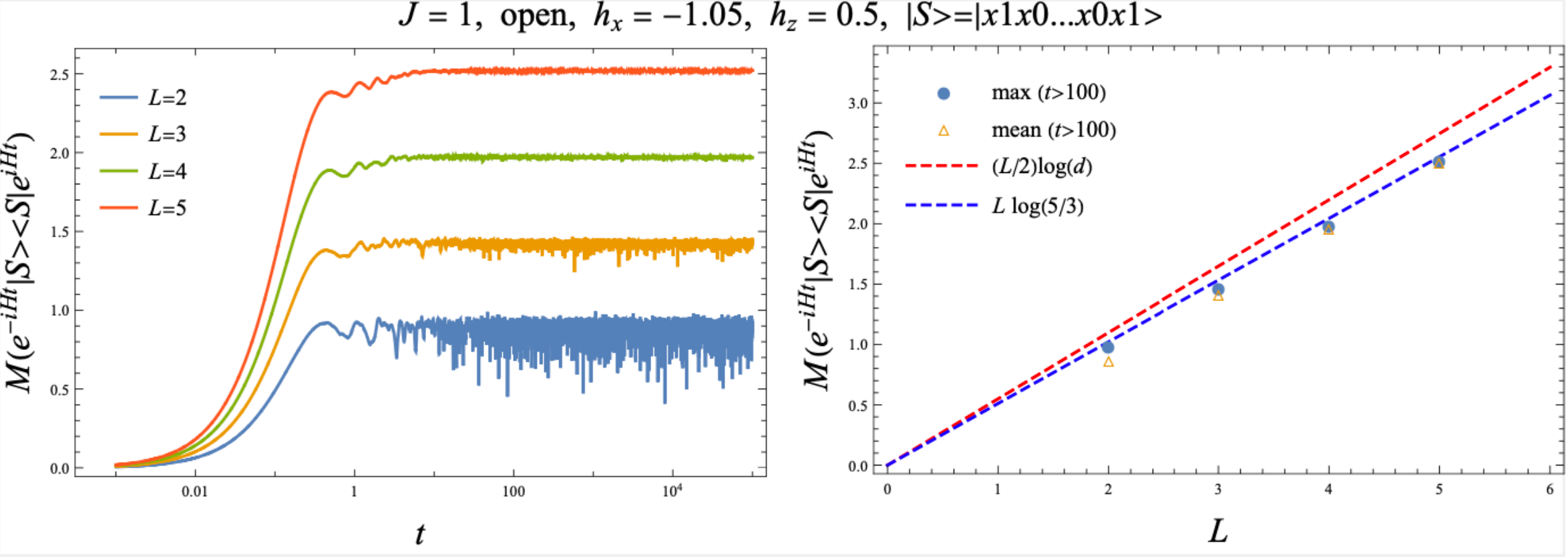}\\
\includegraphics[width=12cm]{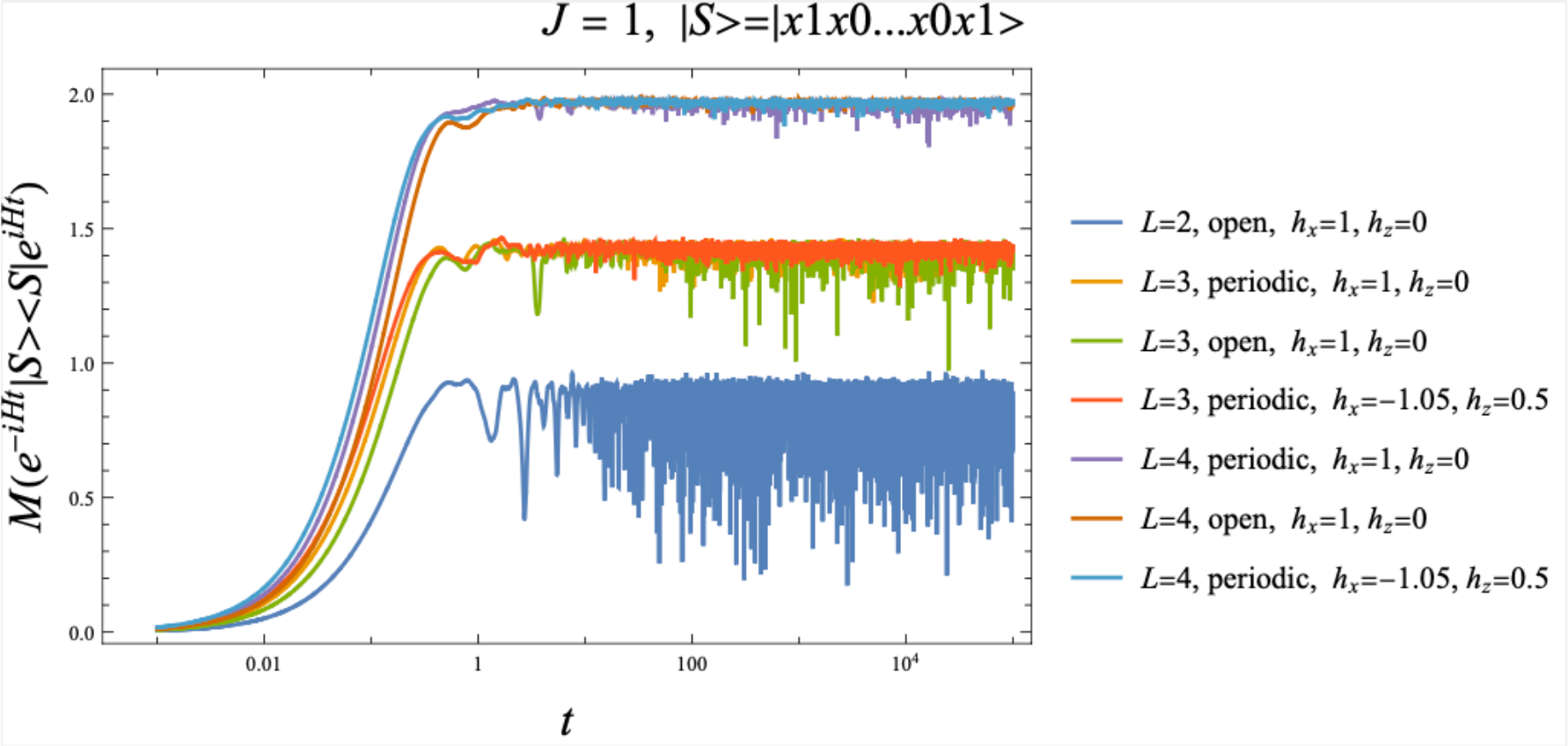}
\caption{Top left: Time evolution of Mana $M(e^{-iHt}|S\rangle\langle S|e^{iHt})$ with $J=1$, $(h_x,h_z)=(-1.05,0.5)$ and $|S\rangle=|x=\omega^0\rangle\otimes |x=\omega\rangle^{\otimes (L-2)}\otimes |x=\omega^0\rangle$;
Top right: Maximum value and the time average of $M(e^{-iHt}|S\rangle\langle S|e^{iHt})$ for $100<t<10^5$.
Bottom: Time evolution of Mana for $h_x=-1.05, h_z=0.5$ with periodic boundary condition and for $h_x=1, h_z=0$ with the open/periodic boundary condition.
}
\label{ManaChaos}
\end{center}
\end{figure}

\begin{figure}
\begin{center}
\includegraphics[width=8cm]{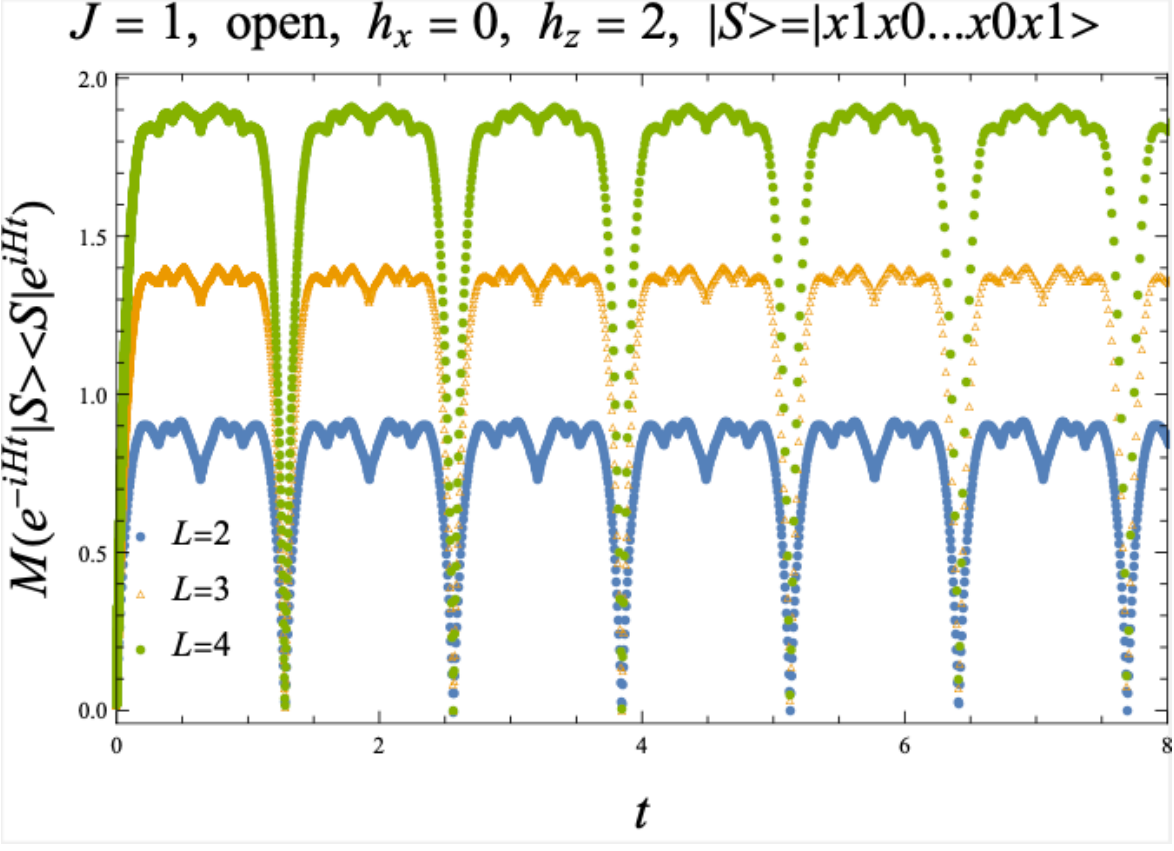}\quad
\includegraphics[width=8cm]{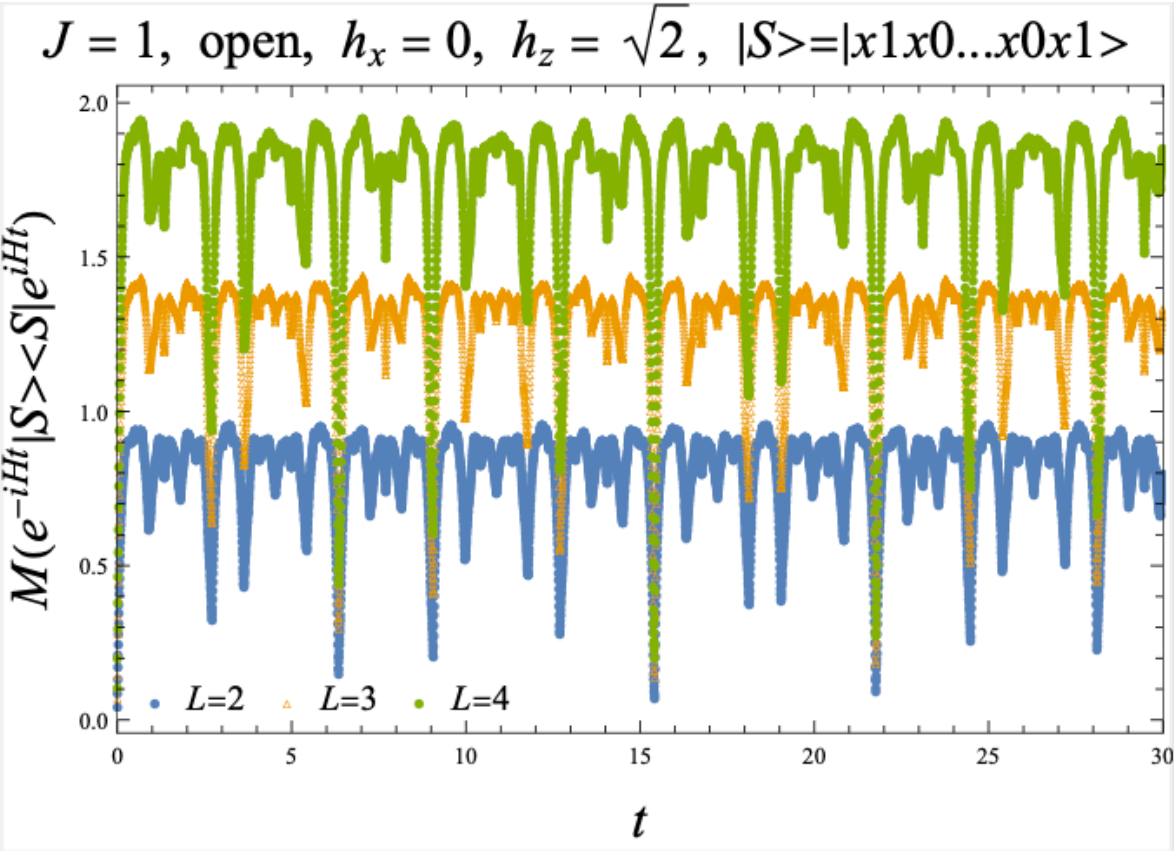}\\
\includegraphics[width=8cm]{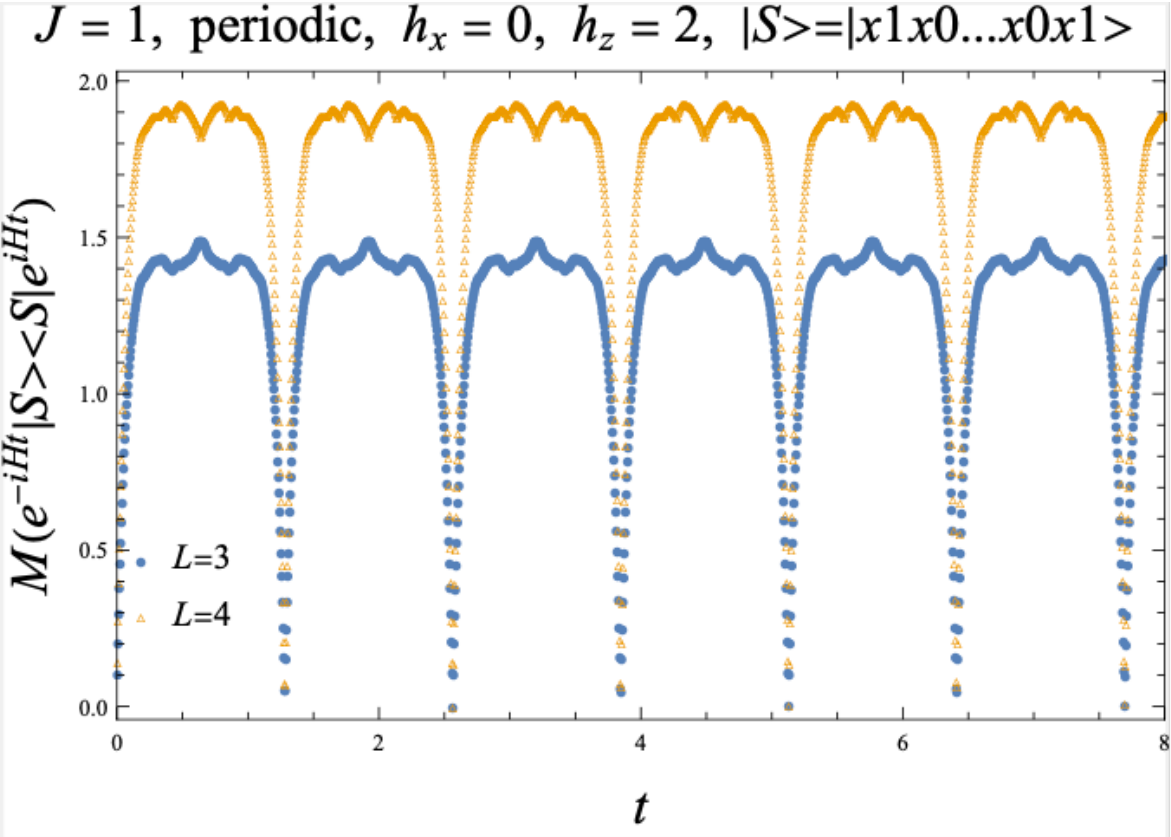}\quad
\includegraphics[width=8cm]{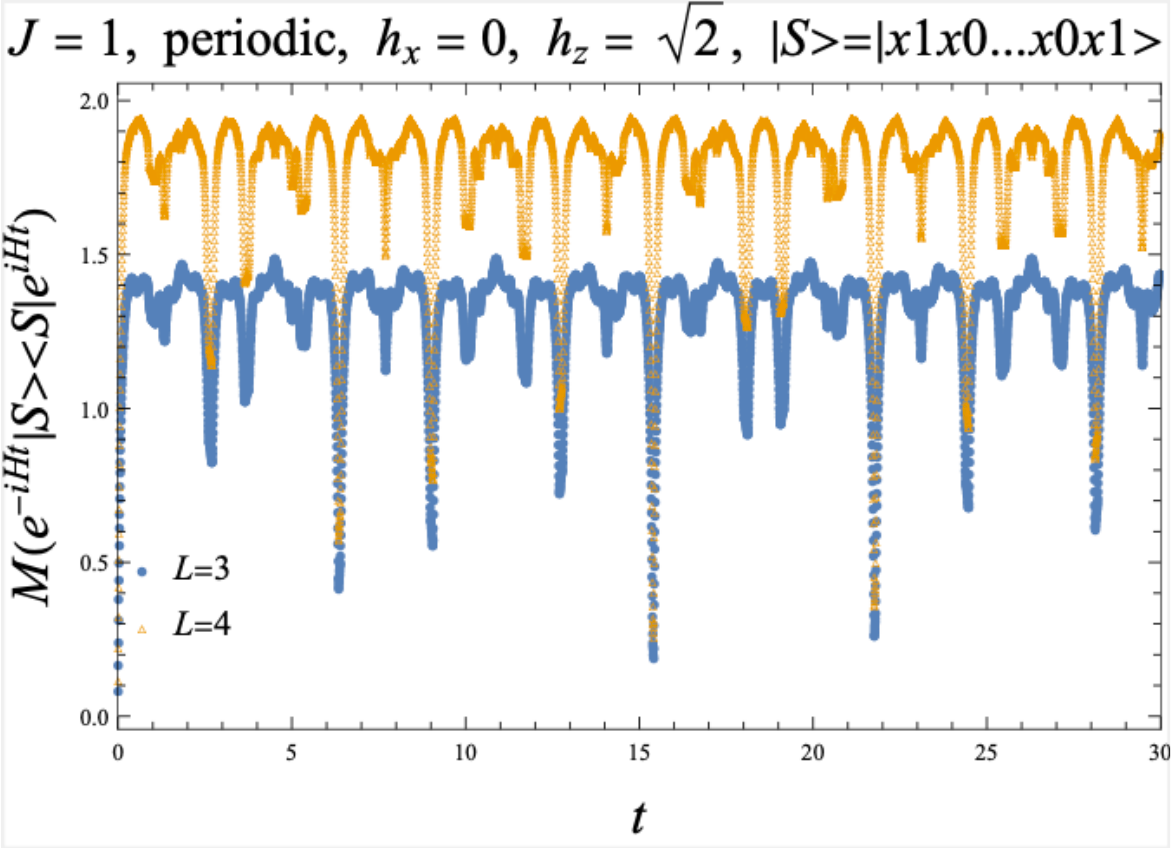}
\caption{Top left: time evolution of Mana $M(e^{-iHt}|S\rangle\langle S|e^{iHt})$ with $J=1$, $(h_x,h_z)=(0,2)$ and $|S\rangle=|x=\omega^0\rangle\otimes |x=\omega\rangle^{\otimes (L-2)}\otimes |x=\omega^0\rangle$;
Top right: the same plot for $(h_x,h_z)=(0,\sqrt{2})$.
Bottom left/Bottom right: The same plots for the periodic boundary condition.
}
\label{ManaInt}
\end{center}
\end{figure}

In figures \ref{ManaChaos} and \ref{ManaInt}, we display the results of Mana for the generalized higher spin Ising model \eqref{HhigherspinIsing} with $J=1$ (i.e.~qudits with $d=3$) in the chaotic regime $(h_x,h_z)=(-1.05,0.5)$ and $(1,0)$ and in the integrable regime $(h_x,h_z)=(0,2)$ and $(0,\sqrt{2})$.

In the chaotic regime $(h_x,h_z)=(-1.05,0.5),(1,0)$, we found that Mana increases monotonically at early times, and then oscillates erratically around some non-zero value.
In particular, we observed that the minima of the oscillation are well separated from zero, the initial value of Mana.
We confirmed that these behaviors are universal for a generic choice of $|S\rangle\in\{|S\rangle\}$, and also do not depend on whether the boundary condition is open/periodic (see appendix \ref{statedependenceofManaandRoM}).
As the number of sites $L$ increases, the maximum value of the oscillation also increases while the magnitude of the error becomes more and more suppressed.
We also observed that the maximum value is very close to $L\log(5/3)$, the lower bound on the optimal upper bound $M_0(3,L)$ on Mana of the pure states \eqref{window}.
Since the window for $M_0(3,L)$ \eqref{window} is narrow, this observation would also be rephrased that the late-time maximum value almost saturates the actual optimal upper bound, although the precise value of $M_0(3,L)$ is still undetermined.

 Here purely based on the results of the numerical analysis, we proposed for finite $d$ ($d=3$) that the late-time maximum of Mana almost saturates the optimal upper bound. Note that for a chaotic system with $d\gg 1$ that enjoys Haar randomness, there is a different argument to justify the same statement as follows.
Let us assume that Mana becomes almost independent of the choice of the initial state at late times as we observed for $d=3$.
Then we would be able to evaluate the late-time Mana of a single state as an average over Haar random ensemble
\begin{align}
M(e^{-iHt}|\psi\rangle\langle\psi |e^{iHt})\approx \langle M(|\psi\rangle\langle\psi|)\rangle_{\text{Haar}},
\label{haar}
\end{align}
where $\langle f(|\psi\rangle)\rangle_{\text{Haar}}$ is defined as ${\cal N}\int_{\sum_{i}|\alpha_{i}|^2=1}\prod_i\frac{i|d\alpha_i|^2}{2}f(|\psi\rangle=\sum_{i}\alpha_i|i\rangle)$ with $\{|i\rangle\}$ an orthonormal basis and ${\cal N}$ the normalization constant.
The right-hand side of \eqref{haar} was calculated for $d\gg 1$ \cite{2020arXiv201113937W} and was found to be $\langle M(|\psi\rangle\langle\psi|)\rangle_{\text{Haar}}\approx (L/2)(\log d-\log\sqrt{\pi/2})$.
Since this can be approximated in the large $d$ limit by $(L/2)\log d$, i.e., the largest possible value for the optimal upper bound $M_0(d,L)$ \eqref{Jensenpure}, we conclude that the late-time Mana saturates $M_0(d,L)$.

In the integrable regime $(h_x,h_z)=(0,2)$, we found that Mana behaves periodically in time and returns to the initial value repeatedly. 
We also studied integrable regime with irrational coupling $(h_x,h_z)=(0,\sqrt{2})$ and found a similar behavior, although its time evolution is not completely periodic in this case.
These results suggest that one can distinguish chaotic systems from integrable systems using Mana.

\subsection{Robustness of Magic\label{time_evo_RoM}}
In the previous section, we found that under chaotic dynamics, the state evolves from a stabilizer state, which can be efficiently simulated on a classical computer, to a state with an almost maximal value of Mana.
This fact is independent of the boundary conditions, and suggests that under chaotic dynamics, except for exceptional states such as energy eigenstates, almost all states evolve to the most difficult state to simulate on a classical computer.
To make sure that this correctly reflects the property of magic in the chaotic regime, rather than the particular property of Mana, we will study another magic measure, Robustness of Magic (RoM).

Let us consider the time evolution of RoM \eqref{211112RoM} of the state $\rho=e^{-iHt}|S\rangle\langle S|e^{iHt}$.
The definition of the RoM involves the optimization over the space spanned by all of the stabilizer pure states.
Since the number of the stabilizer pure states increases quickly with respect to the number of sites $L$ and the dimension of the single-site Hilbert space $d$ as $|\{|S\rangle\}|=d^L\prod_{n=1}^L(d^n+1)$ \eqref{211221numberofstabilizerpurestates}, the computation of RoM is difficult compared to  Mana for the same values of  $d$ and $L$.
On the other hand, one advantage of studying RoM is that it can also be defined when $d$ is even in contrast to Mana.

\begin{figure}
\begin{center}
\includegraphics[width=8cm]{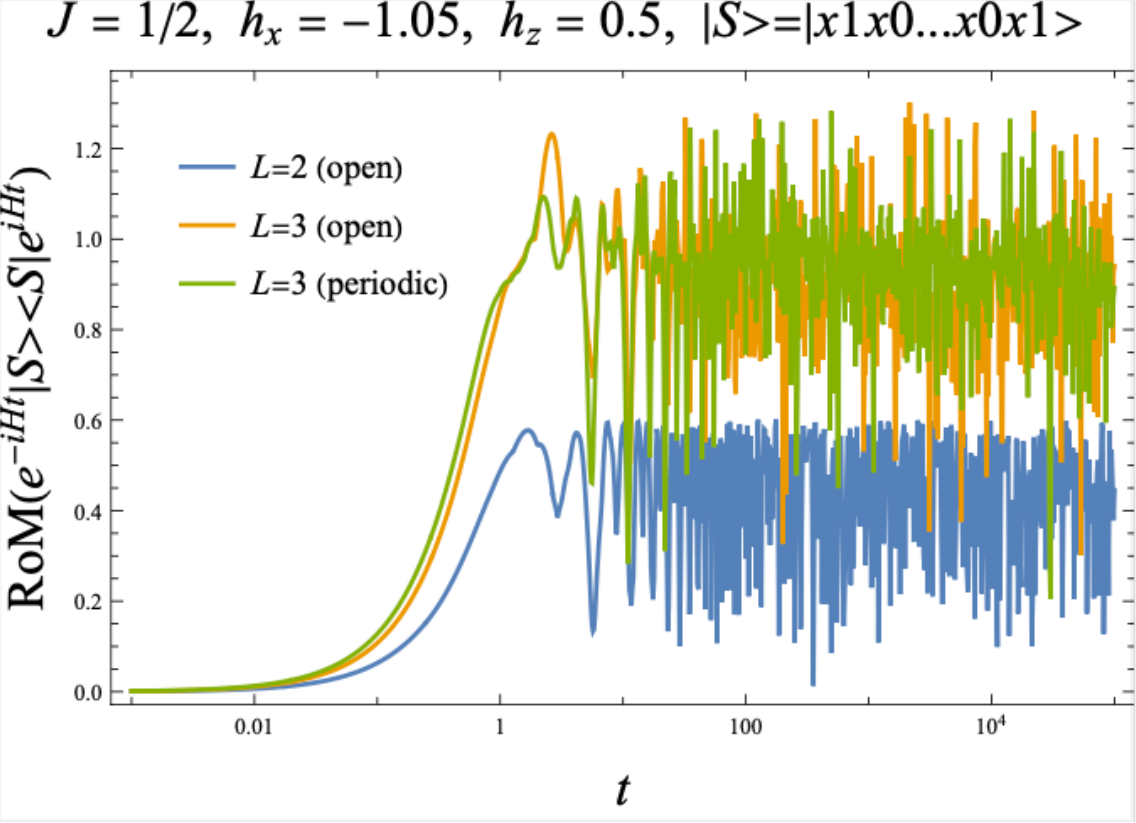}
\includegraphics[width=8cm]{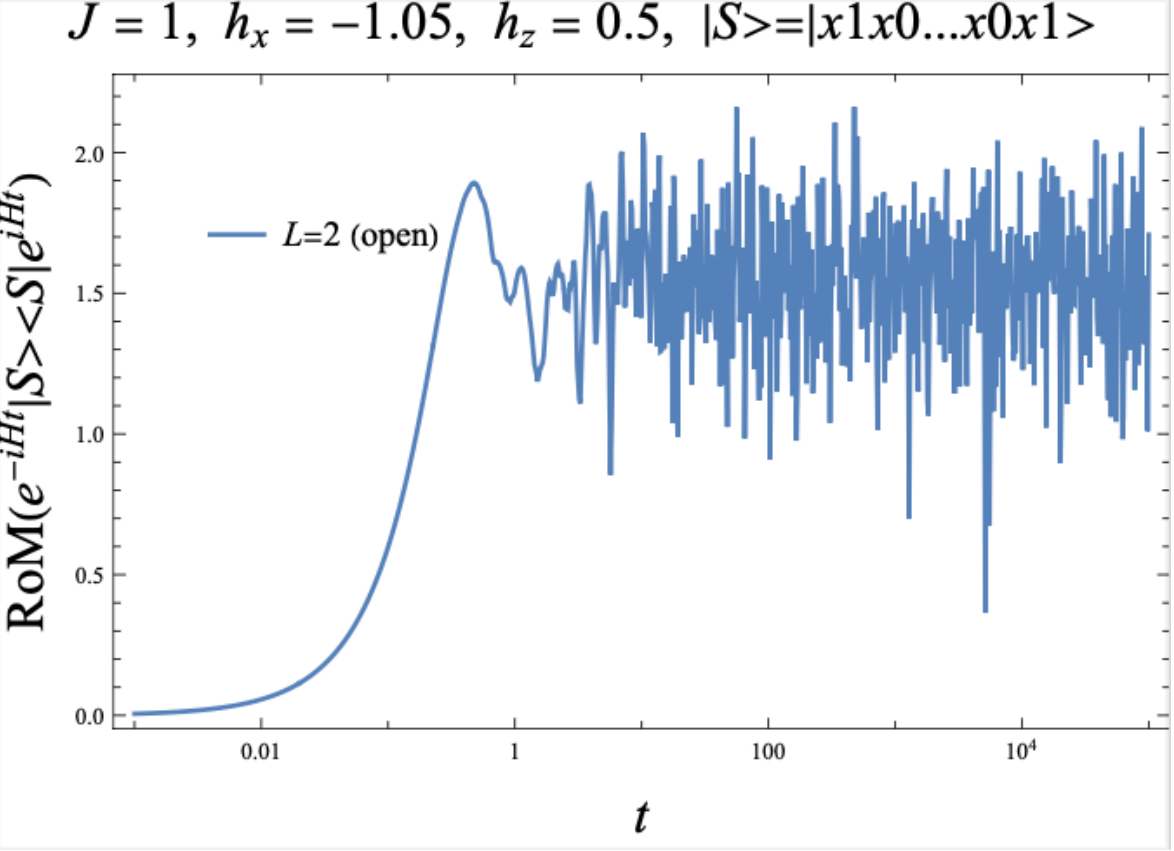}
\caption{
In both the right and left panels, we show the time evolution of the RoM in the chaotic region. In the left panel, we take the parameters $(J,h_x,h_z)$ to be $(1/2,-1.05,0.5)$, while in the right panel, we take the parameters $(J,h_x,h_z)$ to be $(1,-1.05,0.5)$.
}
\label{211127_RoMchaotic}
\end{center}
\end{figure}

\begin{figure}
\begin{center}
\includegraphics[width=8cm]{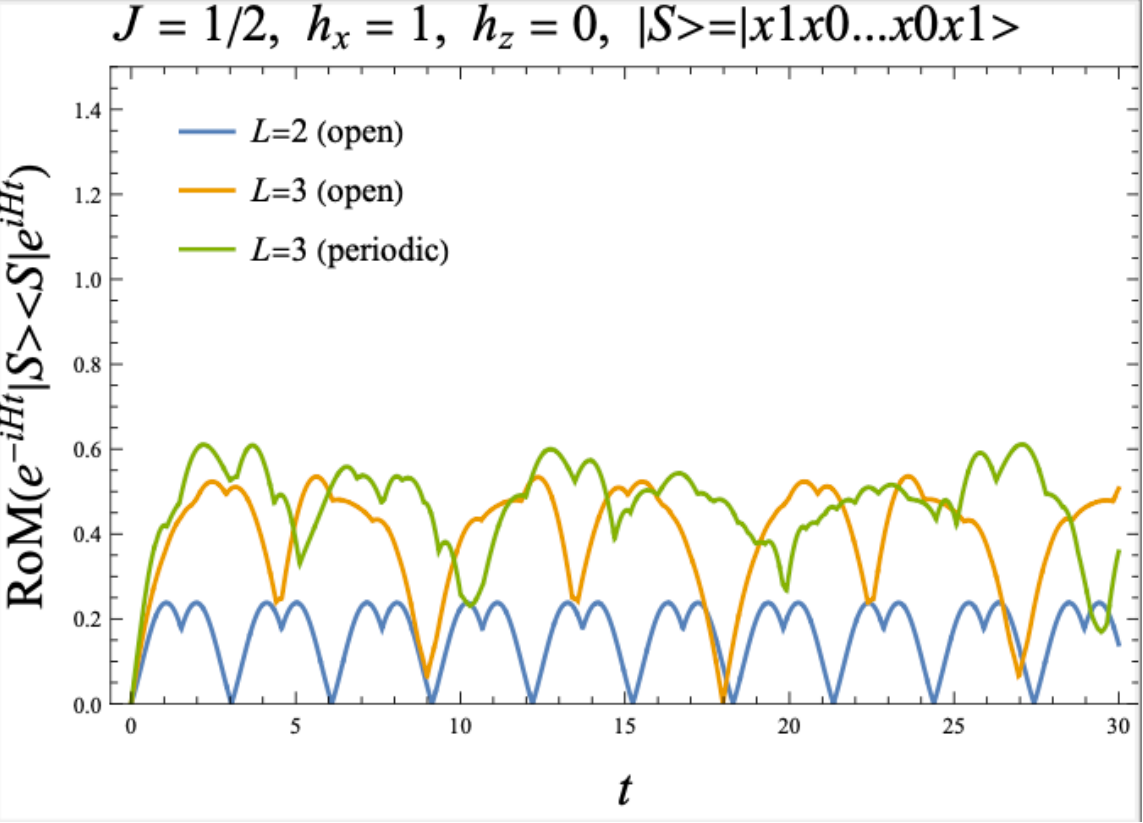}
\includegraphics[width=8cm]{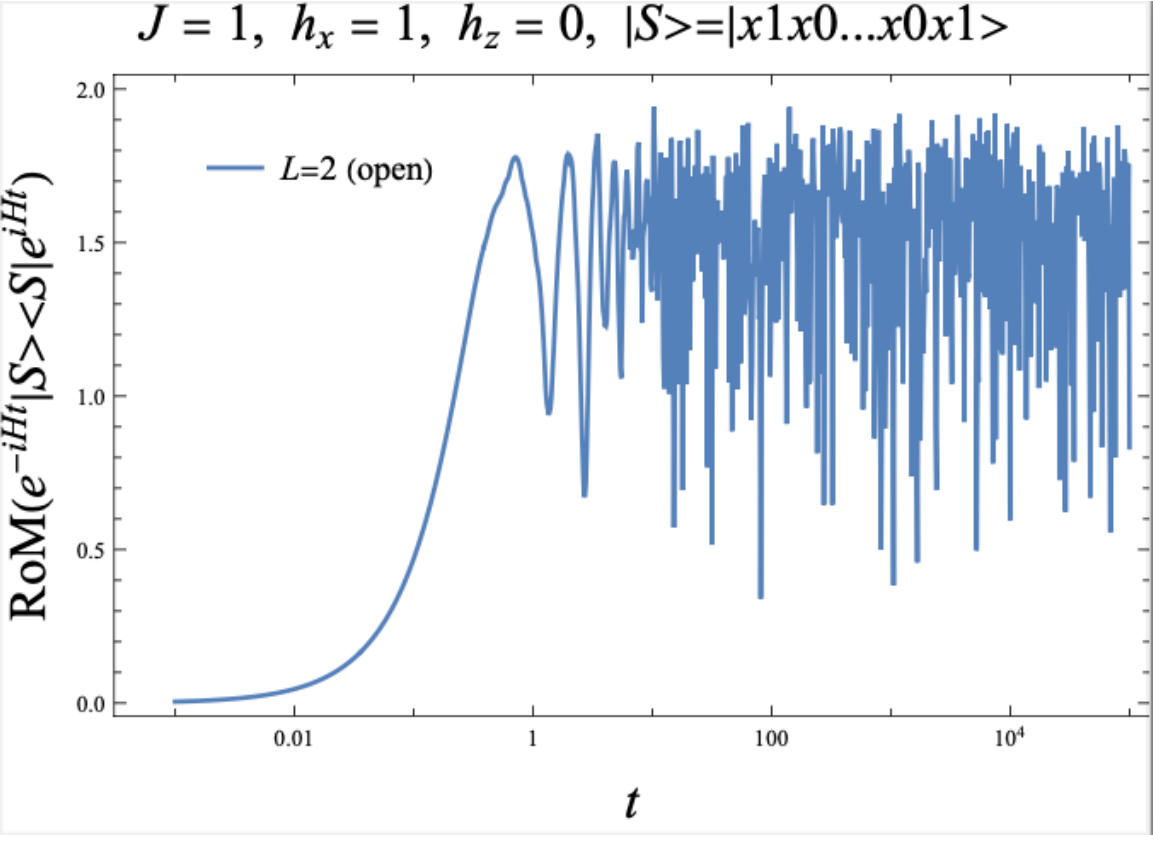}
\caption{
The evolution of RoM in the  higher-spin generalized Ising model with $(J,h_x,h_z)=(1/2,1,0)$ (Left) and $(J,h_x,h_z)=(1,1,0)$ (Right).
}
\label{211213_RoMhx1hz0}
\end{center}
\end{figure}

\begin{figure}
\begin{center}
\includegraphics[width=8cm]{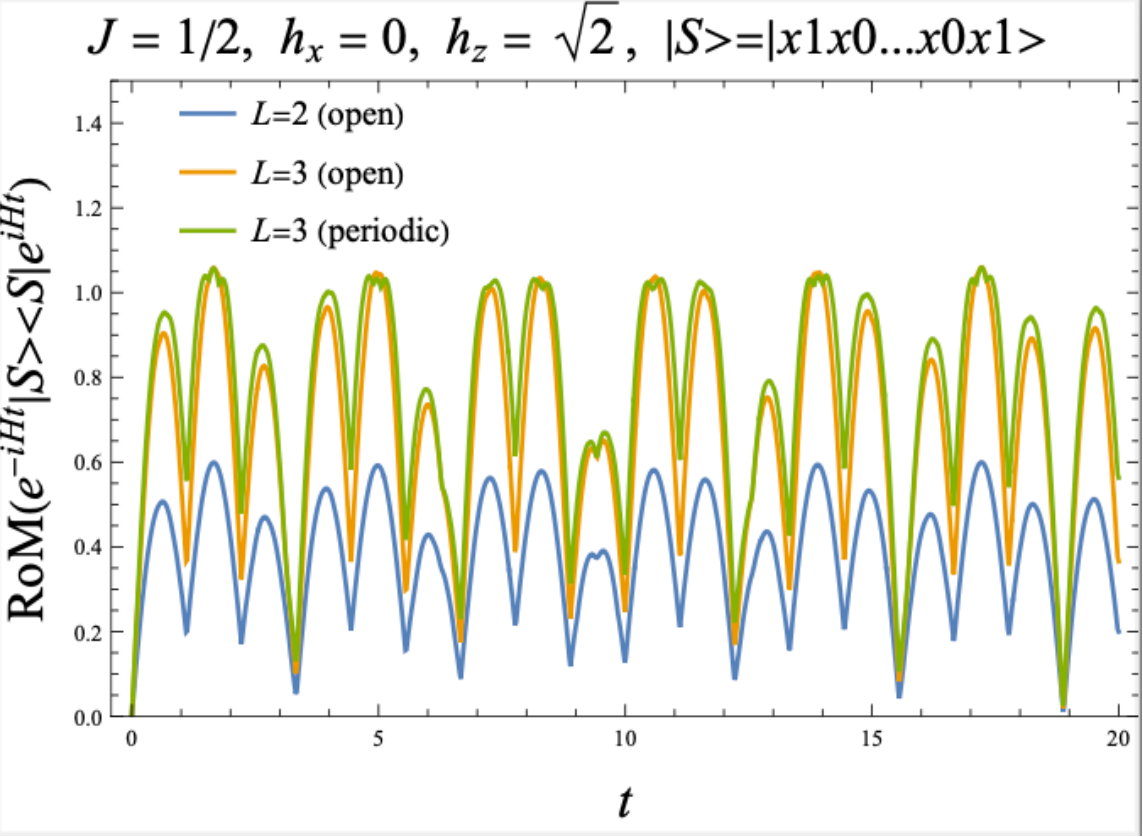}
\includegraphics[width=8cm]{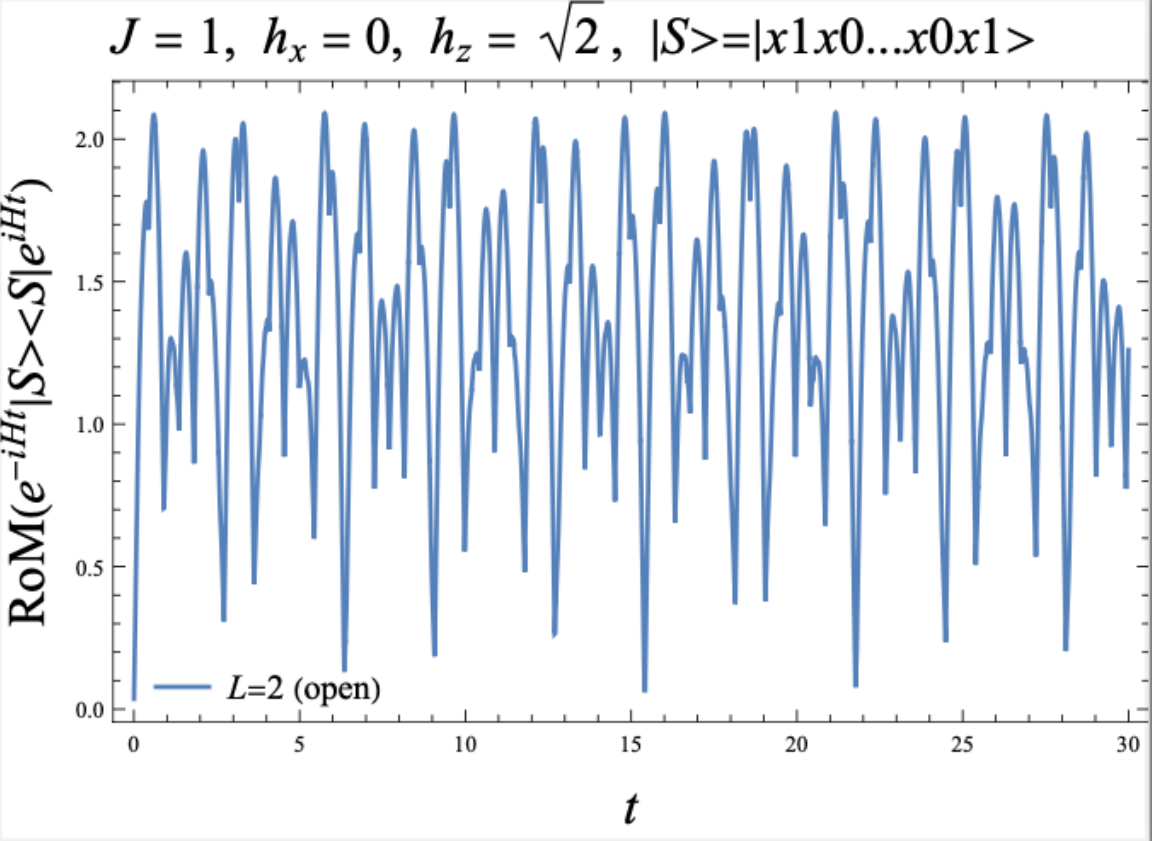}
\caption{
The evolution of RoM in generalized higher-spin Ising model in the integrable regime, $(h_x,h_z)=(0,\sqrt{2})$. In both the right and left panels, we show the time evolution of the RoM in the integrable region. In the left panel, we take the parameters $(J,h_x,h_z)$ to be $(1/2,0,\sqrt{2})$, while in the right panel, we take the parameters $(J,h_x,h_z)$ to be $(1,0,\sqrt{2})$.
}
\label{211127_RoMintegrable}
\end{center}
\end{figure}

In figures \ref{211127_RoMchaotic}, \ref{211213_RoMhx1hz0} and \ref{211127_RoMintegrable}, we displayed the results of the higher-spin generalized Ising model \eqref{HhigherspinIsing} for $(h_x,h_z)=(-1.05,0.5), (1,0), (0,\sqrt{2})$ with $J=1/2,1$.
We found that, at least for $(h_x,h_z)=(-1.05,0.5)$ (chaotic regime) and $(h_x,h_z)=(0,\sqrt{2})$ (integrable regime), the time evolution of RoM exhibits similar behaviors as Mana:
\begin{itemize}
\item For the chaotic case, RoM monotonically increases at early times, and it oscillates between the maximum value and the non-zero minimum value at late times.
The maximal value of RoM becomes larger as $L$ increases.
\item For the integrable case with $(h_x,h_z)=(0,\sqrt{2})$, RoM repeatedly comes close to zero in a short time.
\item RoM in the chaotic regime exhibits the same behavior for the open and the periodic boundary conditions.
We also studied the other choices of the initial stabilizer state and found similar results except for some exceptional cases. One case is that the initial state is an eigenstate of $H$, which does not evolve in time and always gives zero RoM. Another case is when the stabilizer pure state belongs to a sector protected by symmetry that consists only of two energy eigenstates. In this case, RoM behaves periodically with the period given by the difference of the two energy eigenvalues (the same phenomenon can also be seen for Mana).
\end{itemize}

For $(J,h_x,h_z)=(1,1,0)$, we found a qualitatively similar behavior of RoM in the case with $(J,h_x,h_z)=(1,-1.05,0.5)$, which is consistent with the fact that the system is chaotic in both parameter regime.
On the other hand, for $(J,h_x,h_z)=(1/2,1,0)$ with $L=3$ and the periodic boundary condition, we did not find a significant difference in the behavior of RoM from that for $(J,h_x,h_z)=(1/2,-1.05,0.5)$, although the system is integrable in the former case.
We expect that this is a finite $L$ artifact. This is because we also found that the behavior of RoM for $(J,h_x,h_z)=(1/2,1,0)$ and $L=3$ with the periodic boundary condition is different from that for $(J,h_x,h_z)=(1/2,1,0)$ and $L=3$ with the open boundary condition, where RoM approaches zero as in the other integrable cases.

\subsection{Observations and possible interpretations\label{Observe}}
We have seen the time dependence of Mana and RoM in the chaotic and integrable regimes in the higher-spin generalized Ising model. Both quantities behave similarly, suggesting that they correctly capture how magic, the difficulty for a classical computer to simulate the state, evolves in time. While we only computed them in the systems with such small degrees of
freedom that there would be some subtleties in the interpretation of our results. Nevertheless, in this section, we will try to describe what can be observed from the numerical plots shown above and some possible interpretations that can be read from them. \\
{\bf Observations}\\
Here we briefly summarize some observations for the plots displayed in the previous section.
\begin{itemize}
    \item {\bf Chaos and Magic}\\
    In the previous section, we numerically studied the time dependence of Mana and RoM in the chaotic and the integrable regimes of the higher-spin generalized Ising model. We observed that the curves showing the time dependence of Mana and RoM in the chaotic regime initially rise rapidly, then gradually slow down with time, and finally saturate to certain values. They
 behave similarly independently of the choices of the initial states and the boundary conditions except for some special cases.
In particular, at late times, Mana stops varying once it reaches the maximum that scales linearly in the number of sites with the coefficient almost coinciding with the optimal upper bound on Mana for a single qudit system.
In other words, at late times, Mana almost saturates the optimal upper bound for the full $L$ site qudit system.

 On the other hand,  Mana and RoM in the integrable regime behave quite differently from the chaotic regime. They behave periodically with time. They repeatedly transition between the states with large values and the ones with small values of Mana and RoM and never converges to certain values. Their behavior strongly depends on the choices of the initial state and the boundary condition. The typical minimum values of
Mana and RoM at sufficiently late times are smaller than the chaotic regime.
\end{itemize}
{\bf Possible interpretations}
\begin{itemize}
    \item {\bf Maximally Magical State as Typical State}\\    In a chaotic system, one can expect that almost all the states reach the typical states, which share the common features at least probed by some restricted set of the physical observables. As we have seen in the above section, Mana and RoM in the chaotic regime behave independently of the  choices of the initial states and the boundary conditions. In this sense, at late times, almost all the states approach the typical states, who share the common feature from the perspective of magic.
        Moreover, at late times, Mana almost saturates the optimal upper bound.
    This indicates the following statement:
    
    \fbox{
   \begin{minipage}[c]{15cm}
   Typical states in the chaotic systems are almost maximally magical, at least, as probed by Mana and RoM. \end{minipage}}
   
  This is one of the main findings of this paper.
   As described above, we also observed in the integrable regime, Mana and RoM behave periodically. This indicates that the
states never settle down to the maximally magical state in contrast to the chaotic regime,
periodically transitioning between highly magical and less magical states.

     \item {\bf Growth of the Number of Magic Gates}\\
     In the plots described above, we can observe that at early times, Mana and RoM grow rapidly in time and gradually slow down as time evolves. There might be some subtleties and finite $L$ artifacts in our analyses, but if it is the case, it suggests that the number of magic gates required to construct a new time-evolved quantum state already almost saturates in the early stage of time evolution. 

\end{itemize}

\section{Magic in Holography \label{DaHM}}
As we described in the introduction, our original motivation was to understand the role of ``quantumness'' of the boundary non-gravitational system in the emergence of classical spacetime geometry in the context of holography. The holographic duals to the entanglement entropy and computational complexity were proposed in the literature. They revealed intriguing connections between the quantum nature of the conformal field theories and the properties of spacetime geometry in the anti-de Sitter space.

In the previous section, we numerically analyzed Mana and RoM in the chaotic system with small degrees of freedom, which does not have classical gravity dual. We expect the behaviors of Mana and RoM contain some artifacts due to the smallness in the degrees of freedom of the system. Therefore, it is difficult to accurately estimate their behavior in the holographic systems, i.e., chaotic systems with large degrees of freedom. Here, we will only mention a few possibilities\footnote{Some of the readers might be concerned about whether Mana and RoM can be defined in the holographic systems, i.e., continuum field theories without a clear notion of ``qubits''. In this section, we leave this issue as an interesting future direction and simply assume that Mana and RoM are well defined in the holographic systems.}.
To make the discussion concrete, let us consider the holographic conformal field theories and focus on Mana in the so-called  (quenched) thermofield double state: 
\begin{equation} \label{quenchedTFD2}
    \left|{\rm TFD}(t)\right\rangle = \mathcal{N} e^{\frac{-(it+\beta)(H_1+H_2)}{2}}\sum_{E_a} \left|E_a\right \rangle_1\otimes \left|E_a\right \rangle_2,
\end{equation}
defined in the doubled Hilbert space $\mathcal{H}=\mathcal{H}_1 \otimes \mathcal{H}_2$. Here $\left|E_a\right\rangle_i $ is an eigenstate of Hamiltonian $H_i$ with eigenvalue $E_a$ in Hilbert space $\mathcal{H}_i$ and $\beta$ is the inverse temperature, $\mathcal{N}$ is a normalization constant $
    \mathcal{N}^2=\left(\sum_{E_a}e^{-\beta E_a}\right)^{-1}$.\\
{\bf Upper Bound on Mana}\\
Let us consider a chaotic discrete system with dimension $d$ of local Hilbert space at each lattice site and the system size $2L$.
If we assume\footnote{
This assumption is justified once we identify the qudit system with a holographic two-dimensional conformal field theory with central charge $c\gg 1$ at finite temperature $\beta$ (normalized by the UV cutoff scale) through $d=e^{\pi c/(3\beta)}$ by identifying the thermal entropy of both systems as $L\log d=\frac{c\pi L}{3\beta}$ ($L$ on the right-hand side is the one-dimensional volume divided by the UV cutoff, which corresponds to the number of lattice sites).
}
$d\gg 1$, we find, as commented in section \ref{time_evo_mana}, that Mana of any state at late times would almost saturate Jensen's inequality: $M(t)\approx L\log d$.
Notice that the right-hand side gives the thermal entropy $S_{\text{thermal}}$ of the system.
We expect that Mana also satisfies this bound in the holographic systems. In particular in a holographic two-dimensional conformal field theory, we have
\begin{equation} \label{upper_bound}
M(t) \le S_{\text{thermal}}=\frac{c \pi L}{3\beta}
\end{equation}
with the central charge $c$ of the conformal field theory, which would be saturated at late times.\\
{\bf Growth of Mana}\\
Growth of Mana will give us important information about  geometrical structure in the anti-de Sitter space dual to the state (\ref{quenchedTFD2}). It is conjectured that the gravity dual to the state (\ref{quenchedTFD2}) is a wormhole connecting two outside regions, each of which is described by $\mathcal{H}_i$ of the dual quantum system. Therefore, if Mana is properly defined in the holographic systems, its growth can capture the time dependence of some geometrical property of the wormhole.

We raise several possibilities in the growth of Mana in the holographic systems:
\\ {\bf 1. Mana Grows Fast at Early Times, Gradually Slows Down}\\
In the numerical analyses of the higher-spin generalized Ising model, we observed that Mana (and RoM) grow rapidly at early times and gradually slow down as time evolves. There are ambiguities caused by the smallness in the degrees of freedom of the system, it is unclear how the time dependence of Mana in the holographic system would be. One of the most natural guess is that it just shows the same behavior observed in the higher-spin generalized Ising model: Mana firstly grows rapidly, gradually slowing down, and asymptotically approaches its maximal value.
    
    One of the most simple characteristics of the wormhole geometry dual to the thermofield double state (\ref{quenchedTFD2}) is its size, which grows linearly in time.  This time dependence can be captured by  entanglement entropy at the early stage of the time evolution, and computational complexity even at later stages.

    If Mana in the holographic systems indeed behaves as described above, this is in contrast to the time dependence of the size of the wormhole geometry dual to (\ref{quenchedTFD2}). This suggests that the gravitational counterpart of magic, if it exists, is not simply a measure of the size of the wormhole in contrast to the case of entanglement entropy and computational complexity.
\\ {\bf 2. Mana Grows Linearly in Time until It Saturates the Bound}\\
While the growth of Mana in the higher-spin generalized Ising model shows a non-trivial curve as explained above, in the early time region it can be well approximated by a linear function as any smooth curves can be done so in general. Thus, one can imagine another scenario, where only the linearly growing regime in the system with small degrees of freedom is amplified, and the slowing-down regime is suppressed as we increase the degrees of freedom in the system. If it is the case, Mana in the holographic system grows linearly as $M(t)\sim at +b$ and captures the linear-growth in the size of the wormhole in time up to the time when it saturates the upper bound (\ref{upper_bound}).
In particular, one can estimate that the growth of Mana stops at most polynomial time in $S_{\rm thermal}$, assuming that $a$ and $b$ scales at most in the polynomial of $S_{\rm thermal}$, much earlier than the saturation time of the computational complexity. Let us remind ourselves that while the computational complexity counts all the gates equally, magic only counts the magic gates that cannot be efficiently simulated on a classical computer. Therefore, if Mana follows the linear growth discussed above, it suggests that after a sufficiently long time $t\gtrsim S^n_{\rm thermal}$, the state largely deviates from the one that can be efficiently simulated on a classical computer. Here, $n$ is an integer.
\\{\bf 3. Exponential of Mana Grows Linearly in Time until It Saturates the Bound}\\
The third possibility is that exponential of Mana, not Mana itself, grows linearly in time as $e^{M(t)}\sim at+b$. Assuming that $a$ is at most given by the polynomial of $S_{\rm thermal}$ and $e^{S_{\rm thermal}}\gg b$ in the limit $L\gg \beta$, we roughly estimate the saturation time as
\begin{equation} 
    t_{\text{saturation}} \sim e^{S_{\text{thermal}}}\, .
    \label{211221tsaturationsection52_2}
\end{equation}
This is the same order as the saturation time of the computational complexity.\begin{figure}
\begin{center}
\includegraphics[width=16cm]{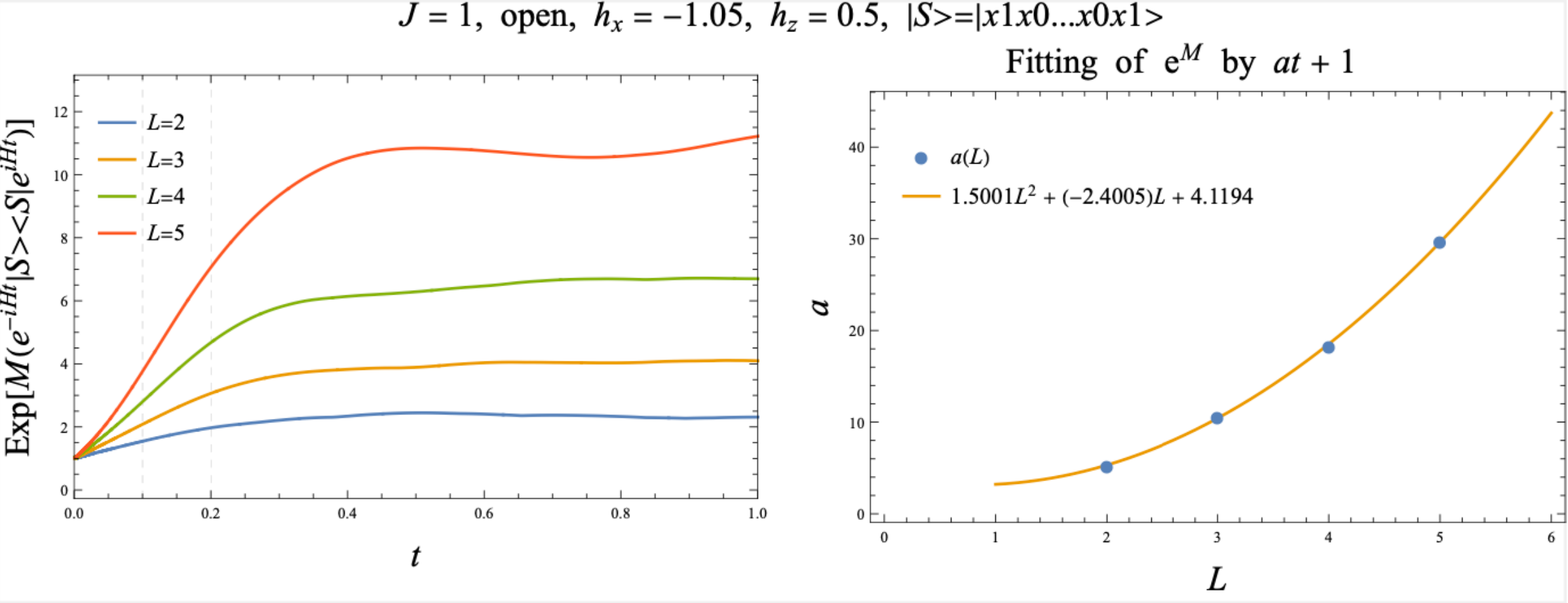}
\caption{
Left: Time evolution of $e^{M}$.
Right: The rate of early-time linear growth of $e^{M}$ obtained by fitting the data in $0.05\le t\le 0.2$.
}
\label{Manaearlytimefitting}
\end{center}
\end{figure}
\begin{figure}
\begin{center}
\includegraphics[width=16cm]{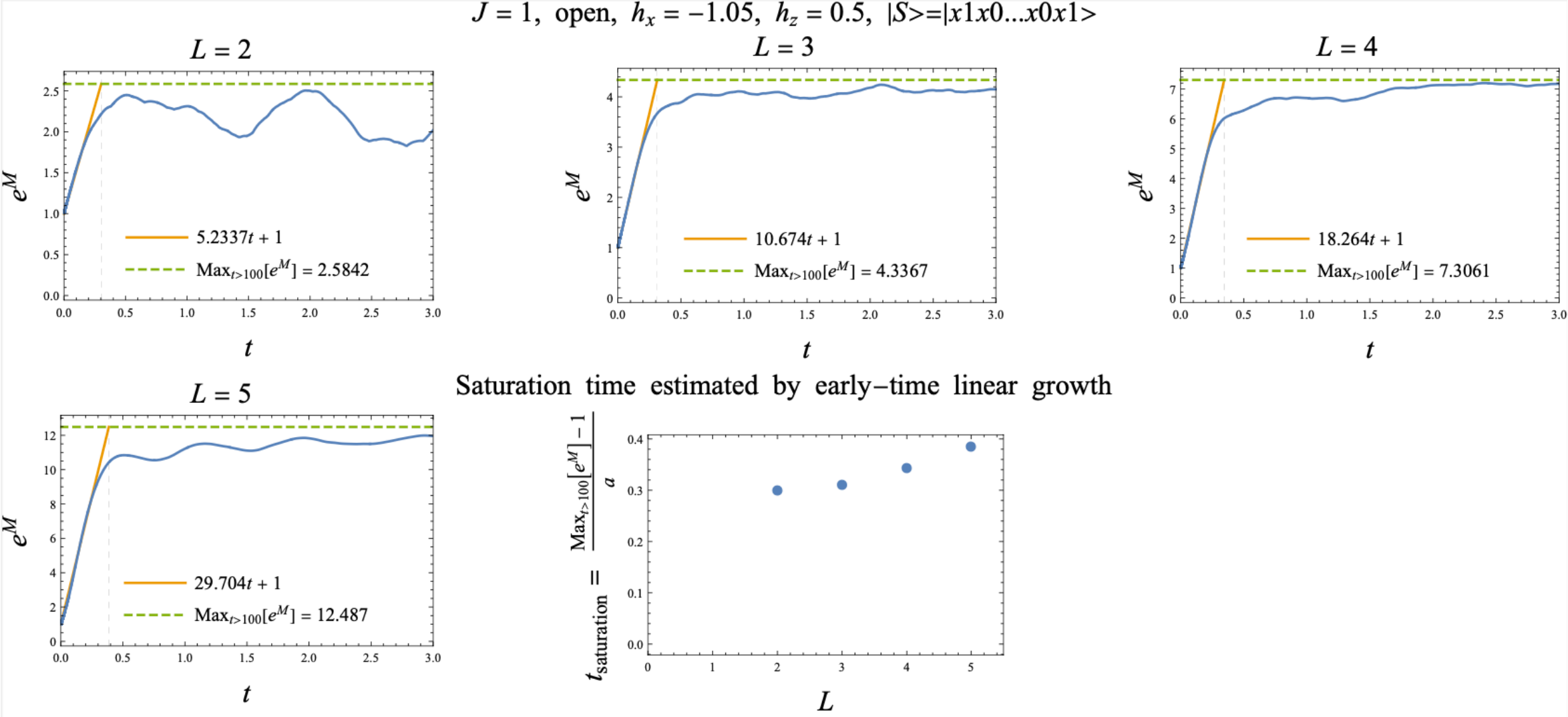}
\caption{
First four panels: comparison between $M$ and its maximum value at late times $(t>100)$ (dashed green line) with the early-time fitting with $e^M=at+1$ (orange line).
Last panel: $L$-dependence of the saturation time estimated by the early-time linear growth (vertical dashed lines in the other panels).
}
\label{Manaestimatedsaturationtime}
\end{center}
\end{figure}
The assumption $e^{M(t)}\approx at+b$ is also supported by the numerical results for the spin system.
Note that the initial states we have used in the numerical analysis are different from the thermofield double state \eqref{quenchedTFD2}.
Nevertheless, here we assume that the behavior of Mana is qualitatively the same also when the initial state is the thermofield double state.
As displayed in figure \ref{Manaearlytimefitting}, we observe that $e^{M(\rho=e^{-iHt}|S\rangle\langle S|e^{iHt})}$ grows linearly at early times, $e^{M}\approx at+1$ with $a$ some constant.
Since at late times, Mana oscillates erratically around some nonzero value, it would be natural to define the saturation time of Mana for the  higher-spin generalized Ising model as
\begin{align} \label{t_sat_ana}
t_{\text{saturation}}=\frac{e^{\text{max}_{t>100}[M(\rho=e^{-iHt}|S\rangle\langle S|e^{iHt})]}-1}{a},
\end{align}
where $\text{max}_{t>10}[M(\rho=e^{-iHt}|S\rangle\langle S|e^{iHt})]$ is the maximum value of Mana at late times.
In figures \ref{Manaearlytimefitting} and \ref{Manaestimatedsaturationtime}, we displayed the results of the fitting at early times and the $L$-dependence of $t_{\text{saturation}}$ obtained.
We found that as the number of sites $L$ increases, both the early-time growth rate $a$ and the saturation time $t_{\text{saturation}}$ increase.
We further observed that as the number of the sites $L$ is increased $a$ scales as $a\sim L^2$.
Combining this result with the observation that the late-time maximum $\text{max}_{t>10}[M(\rho=e^{-iHt}|S\rangle\langle S|e^{iHt})]$ grows linearly in $L$ (see figure \ref{ManaChaos}), it follows that the saturation time scales exponentially in $L$.
Note that there are several subtleties in our analysis in the higher-spin generalized Ising model.
First, since our analysis is limited to the small system size $L\le 5$, it is not clear whether the $L$-dependence of the growth coefficient $a(L)$ is really of polynomial in $L$.
Also, even if the polynomial scaling is correct, the physical interpretation of neither the order of the polynomial nor the coefficients of the polynomial are not clear.
Second, it is not clear whether our result depend on the choice of the initial stabilizer pure state $|S\rangle$:
although we have observed that the late-time maximum of $M(e^{-iHt}|S\rangle\langle S|e^{iHt})$ is almost independent of the choice of $|S\rangle$, for the growth coefficient $a(L)$ it is still not clear, at $L=4$ where we have the results for a moderate variety of $|S\rangle$, whether $a(L)$ tends to be independent of $|S\rangle$ as $L$ increases or not.
It would be an important future direction to clarify these points.\\

\section{Discussions and Future Directions \label{sec_discuss}}
In this work, we studied the time dependence of the two measures of magic, Mana and RoM, in the higher-spin generalized Ising model in the integrable and chaotic regimes.
We chose stabilizer states, which can be efficiently simulated on a classical computer, as initial states, and considered their time evolution under the Hamiltonian of the system.
We found that in the chaotic regime, both Mana and RoM increase monotonically and saturate after a sufficient late time to some non-zero value independent of the choices of an initial state and a boundary condition.
In particular, Mana at late times almost coincides with the optimal upper bound for a single-site system times the number of sites.
This suggests that Mana at late times almost saturates the optimal upper bound on Mana for the entire Hilbert space.
On the other hand, in the integrable regime, Mana and RoM behave periodically in time and the state transitions between low-magic and high-magic states. Our study suggests that magic of a quantum state is one of the key elements for the emergence of chaotic properties.

In this section, we list several further discussions and future directions.

\begin{itemize}
\item 
{\bf Mana in a Thermal State}\\
In this paper, we only focused on magic for a pure state. Here we make a brief comment on speculation on magic for a thermal state. Let us take the system of $L$ sites, which has $d$ degrees of freedom of the Hilbert space on each site, i.e., the dimension of the Hilbert space is $d^L$. We expect that the density matrix of a thermal state approximately takes the form
\be
\rho_{\text{thermal}} \approx \f{1}{d^{L}}\mathbbm{1},
\ee
where $\mathbbm{1}$ is the $d^L \times d^L$ identity matrix.
Computed from this density matrix, the R\'{e}nyi entanglement entropy  becomes independent of the replica index $n$.
Jensen's inequality would suggest that Mana is always zero. It would be intriguing to explore its physical interpretation. \if[0], so that even if the thermal state in this limit is quenched in some way, $M(\rho_{\text{Thermal}})$ is independent of time.
On the other hand, the local Mana defined with respect to the reduced density matrix can have a nontrivial time evolution.\fi
\item {\bf Mana for a Subregion}\\
Mana studied in this paper is defined through the density matrix of a quantum state. We can simply generalize it to Mana for a subregion using the {\it reduced} density matrix.
Let us consider a state defined on $L$ sites and divide the system into $A$ with $l_A$ sites and $B$ with the remaining sites. We also assume that the size dependence of the R\'{e}nyi entanglement entropy follows the Page curve \cite{Page:1993df, Sen:1996ph}.
i.e., the subsystem size dependence of the second R\'{e}nyi entanglement entropy $S^{(2)}_A$ for the region $A$ is given by
\be
\begin{split}
S^{(2)}_A = \begin{cases}
l_A \log{d}& \f{L}{2}>l_A\\
(L-l_A) \log{d} & L>l_A>\f{L}{2} \\
\end{cases}\, .
\end{split}
\ee
This suggests that the upper bound for the subsystem generalization of Mana $\mathcal{M}_A$ defined by the reduced density matrix $\rho_A$ is given by 
\be \label {mana_page}
\begin{split}
\mathcal{M}_A\le \begin{cases}
0 & 0<l_A<\f{L}{2}\\
l_A -\f{L}{2} & \f{L}{2}<l_A<L\, .
\end{cases}
\end{split}
\ee
Especially it would be worth noting that the subregion generalization of Mana $\mathcal{M}_A$ is always $0$ for a small subregion  $0<l<\f{L}{2}$.
It is known that the entanglement entropy in a thermofield double state in a two-dimensional holographic conformal field theory on a compact spacetime \cite{Goto:2021gve} and a chaotic chain \cite{2020arXiv201214609M} follows the Page curve at sufficiently late times, and we expect the subregion generalization of Mana should satisfy (\ref{mana_page})\footnote{Since the time evolution of the R\'{e}nyi entanglement entropy in a two-dimensional holographic conformal field theory depends on the R\'{e}nyi index, a correction to (\ref{mana_page}) may be necessary.}.
\end{itemize}

\subsection*{Future directions}
\begin{itemize}
\item One of the interesting future directions is to find the gravity dual of Mana and RoM.
It would be interesting to find the geometrical meaning of the maximally magical states, which would be realized in the holographic systems.
\item In our paper, we performed numerical computations for Mana and RoM, but there would be some subtleties due to the small system size.
It would be interesting to study them for larger systems. Such direction is important to find the gravity dual of them.
\item It is important to define Mana and RoM in a continuous quantum field theory and study their time dependence.
\item In section \ref{D_o_Mana} we have observed that under chaotic dynamics, any state evolves to some state $|\psi(t)\rangle$ whose Mana is $M(|\psi(t)\rangle\langle\psi(t)|)\approx LM_0(d,1)$, where $L$ is the number of sites, $d$ is the dimension of the single-site Hilbert space, and $M_0(d,1)$ is the optimal upper bound on Mana of the single-site system, at least for $d=3$.
One possible choice of the state to reproduce the same amount of Mana is
\begin{align}
U \bigotimes_{i=1}^L|m_i\rangle,
\label{aproposal}
\end{align}
with each of $|m_i\rangle$ being one of the states in the single-site system whose Mana saturates $M=M_0(d,1)$, and $U$ is some element of the Clifford group $C_d$ of the $L$ site system.
Such single-site states $|m\rangle$ are known concretely for $d=3,5$ \cite{2014NJPh...16a3009V,2020PhRvA.102d2409J}.
It would be interesting to study whether the late-time state can be expressed in the form \eqref{aproposal},\footnote{
Note that the state $\otimes_{i=1}^L|m_i\rangle$ does not have any entanglement.
If a late-time state would be of the form \eqref{aproposal}, the large entanglement of the state, which is a characteristic feature of chaotic dynamics, should be carried by the choice of $U$.
}
or more generally, whether the state at early times where Mana is still growing can also be expressed in the similar form: $U(t)( \otimes_i^{\ell(t)}|m_i\rangle\otimes |S\rangle)$ with $\ell(t)\approx M(|\psi\rangle\langle\psi|)/M_0(d,1)$ and $|S\rangle$ some stabilizer state in the $L-\ell(t)$ site system.
Conversely, it would also be interesting to study whether dynamics under which any state evolves to an almost maximally magical state is always strongly chaotic or not.
\end{itemize}

\section*{Acknowledgements}
We thank Shinsei Ryu, Kotaro Tamaoka, and Tadashi Takayanagi for valuable discussions. 
We thank Tomoyuki Morimae for giving us an introductory seminar on the basics of  magic at YITP, Kyoto University. K.G.~is supported by JSPS Grant-in-Aid for Early-Career Scientists 21K13930. 
M.N.~is supported by JSPS Grant-in-Aid for Early-Career Scientists 19K14724. 
The numerical results were obtained using the high-performance computing facilities provided in SISSA (Ulysses) and in Yukawa Institute for Theoretical Physics (Sushiki server). The sigils for chaos magic used as symbols to indicate the author's affiliations were borrowed from \url{https://en.wikipedia.org/wiki/Chaos_magic#/media/File:Sigil_shoal.svg}.

\appendix

\section{Initial-State Dependence of Mana and RoM}
\label{statedependenceofManaandRoM}
\begin{figure}
\begin{center}
\includegraphics[width=16cm]{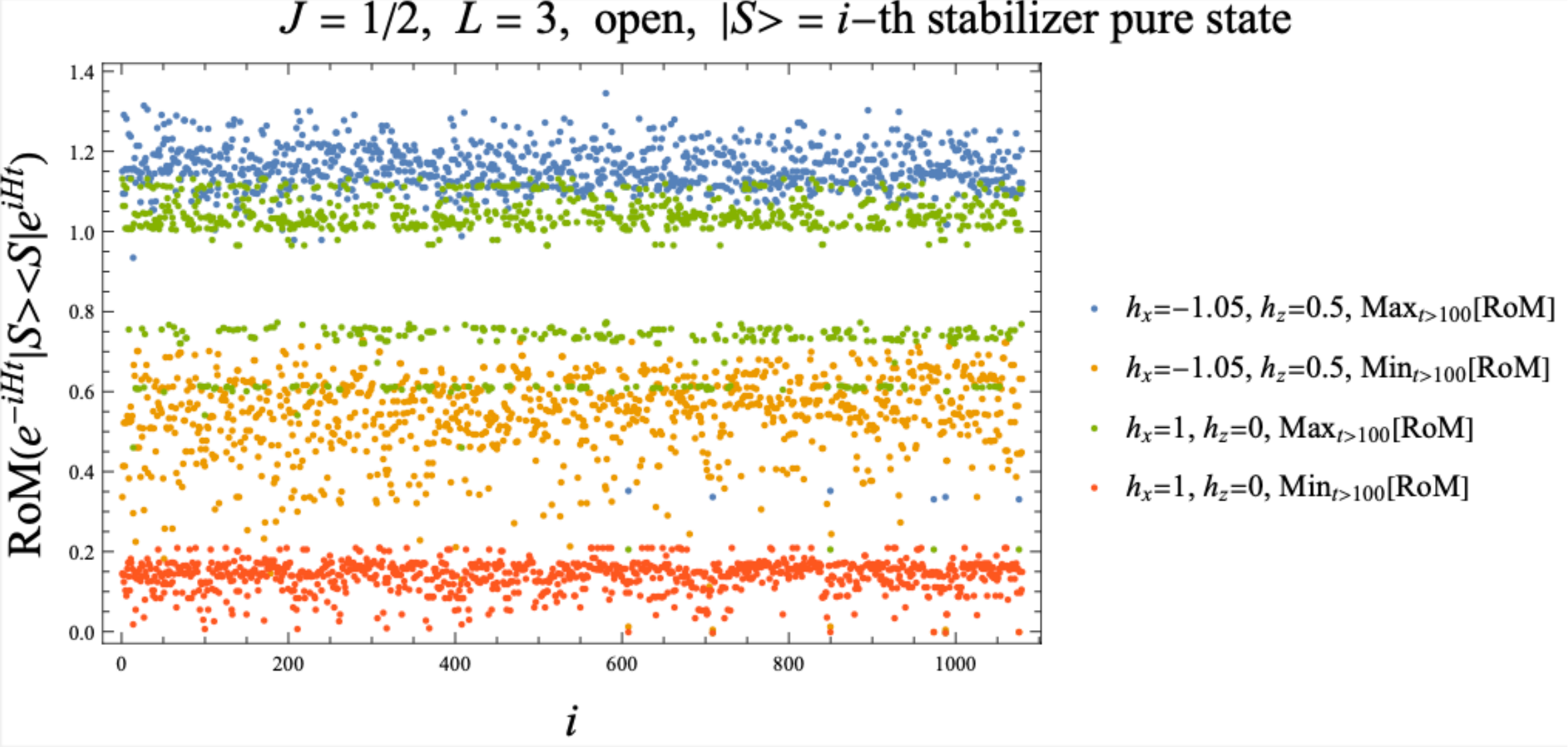}\\
\includegraphics[width=16cm]{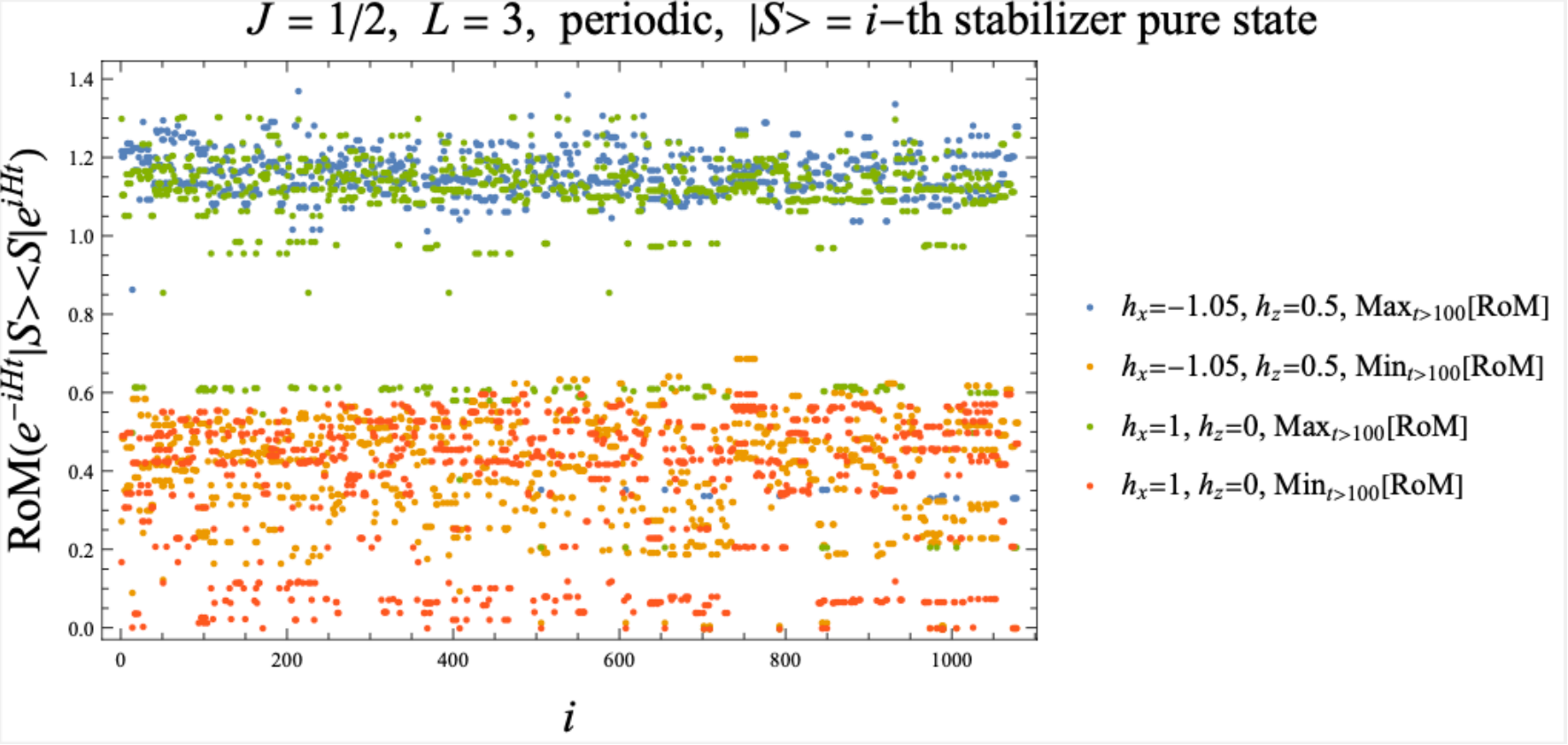}
\caption{Observed maximum and minimum values of $\text{RoM}(e^{-iHt}|S\rangle\langle S|e^{iHt})$ for $L=3$, $(j,h_x,h_z)=(1/2,-1.05,0.5)$ and $(j,h_x,h_z)=(1/2,1,0)$ with $t>100$.
The horizontal axis indicates that the $i$-th stabilizer pure state among all 1080 stabilizer pure states is chosen as the initial state $|S\rangle$.}
\label{RoMmaxminlistj1/2L3}
\end{center}
\end{figure}

\begin{figure}
\begin{center}
\includegraphics[width=16cm]{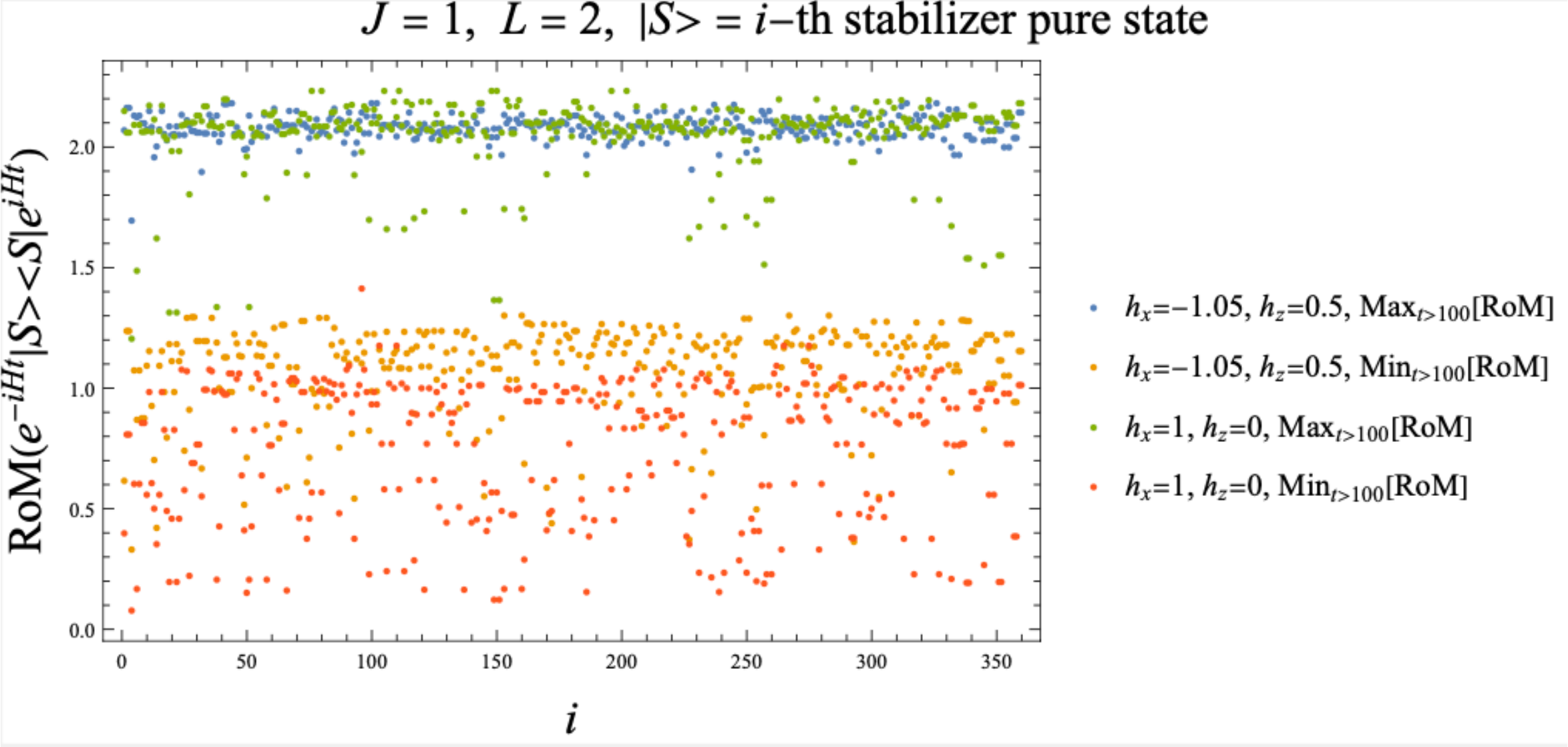}
\caption{Observed maximum and minimum values of $\text{RoM}(e^{-iHt}|S\rangle\langle S|e^{iHt})$ for $L=2$, $(j,h_x,h_z)=(1,-1.05,0.5)$ and $(j,h_x,h_z)=(1,1,0)$ with $t>100$.
The horizontal axis indicates that the $i$-th stabilizer pure state among all 360 stabilizer pure states is chosen as the initial state $|S\rangle$.}
\label{RoMmaxminlistj1L2}
\end{center}
\end{figure}

In this appendix, we display the results on the dependence of Mana $M(\rho)$ and $\text{RoM}(\rho)$ for $\rho=e^{-iHt}|S\rangle\langle S|e^{iHt}$ on the choice of the initial state $|S\rangle$.
For simplicity, we display only the numerically realized values of maximum and minimum at late times ($100<t<10^5$).
We also omit the results for the trivially integrable ($h_x=0$) cases.
In tables \ref{211223_Manalatetimevaluelist1} and \ref{211223_Manalatetimevaluelist2}, we list the results for Mana with $j=1,L=4$, where we have selected only part of the stabilizer states for $|S\rangle$ among the full set of stabilizer pure states, which consists of $7439040$ \eqref{211221numberofstabilizerpurestates}, while in figures \ref{RoMmaxminlistj1/2L3} and \ref{RoMmaxminlistj1L2}, we display the results for RoM with $j=1/2,L=3$ and $j=1,L=2$ for all stabilizer states $|S\rangle$.
\begin{figure}
\begin{center}
\begin{tabular}{|c|c|c|c|c|}
\hline
                  &\multicolumn{2}{|c|}{$h_x=-1.05, h_z=0.5$, open}  &\multicolumn{2}{|c|}{$h_x=-1.05, h_z=0.5$, periodic}\\ \hline
$|S\rangle$       &$\text{Max}_{t>100}[M]$&$\text{Min}_{t>100}[M]$&$\text{Max}_{t>100}[M]$&$\text{Min}_{t>100}[M]$\\ \hline
$|x0x0x0x0\rangle$&$1.94848$              &$1.5446$               &$1.9677$               &$1.35928$\\ \hline
$|x0x1x1x0\rangle$&$1.98872$              &$1.94041$              &$1.98788$              &$1.88011$\\ \hline
$|x0x2x2x0\rangle$&$1.98867$              &$1.92726$              &$1.98738$              &$1.90887$\\ \hline
$|x0z0z0x0\rangle$&$1.98775$              &$1.94232$              &$1.98697$              &$1.93669$\\ \hline
$|x0z1z1x0\rangle$&$1.98389$              &$1.8928$               &$1.98385$              &$1.84364$\\ \hline
$|x0z2z2x0\rangle$&$1.98663$              &$1.8807$               &$1.98492$              &$1.84096$\\ \hline
$|x1x0x0x1\rangle$&$1.98636$              &$1.94038$              &$1.98788$              &$1.88011$\\ \hline
$|x1x1x1x1\rangle$&$1.98384$              &$1.87289$              &$1.9886$               &$1.71676$\\ \hline
$|x1x2x2x1\rangle$&$1.98448$              &$1.92106$              &$1.98814$              &$1.87717$\\ \hline
$|x1z0z0x1\rangle$&$1.97987$              &$1.82827$              &$1.98082$              &$1.85984$\\ \hline
$|x1z1z1x1\rangle$&$1.98799$              &$1.93234$              &$1.99067$              &$1.9296$\\ \hline
$|x1z2z2x1\rangle$&$1.99014$              &$1.91262$              &$1.99007$              &$1.91344$\\ \hline
$|x2x0x0x2\rangle$&$1.98809$              &$1.93112$              &$1.98738$              &$1.90887$\\ \hline
$|x2x1x1x2\rangle$&$1.98586$              &$1.90739$              &$1.98814$              &$1.87717$\\ \hline
$|x2x2x2x2\rangle$&$1.98658$              &$1.83597$              &$1.99177$              &$1.71701$\\ \hline
$|x2z0z0x2\rangle$&$1.9799$               &$1.83753$              &$1.98052$              &$1.85992$\\ \hline
$|x2z1z1x2\rangle$&$1.98999$              &$1.93231$              &$1.9873$               &$1.92717$\\ \hline
$|x2z2z2x2\rangle$&$1.9897$               &$1.9217$               &$1.99008$              &$1.91503$\\ \hline
$|z0x0x0z0\rangle$&$1.98726$              &$1.93525$              &$1.98697$              &$1.93669$\\ \hline
$|z0x1x1z0\rangle$&$1.98632$              &$1.92497$              &$1.98082$              &$1.85984$\\ \hline
$|z0x2x2z0\rangle$&$1.9862$               &$1.91971$              &$1.98052$              &$1.85992$\\ \hline
$|z1x0x0z1\rangle$&$1.98368$              &$1.86783$              &$1.98385$              &$1.84364$\\ \hline
$|z1x1x1z1\rangle$&$1.98769$              &$1.94325$              &$1.99067$              &$1.9296$\\ \hline
$|z1x2x2z1\rangle$&$1.98894$              &$1.94228$              &$1.9873$               &$1.92717$\\ \hline
$|z2x0x0z2\rangle$&$1.97872$              &$1.71465$              &$1.98492$              &$1.84096$\\ \hline
$|z2x1x1z2\rangle$&$1.98956$              &$1.91078$              &$1.99007$              &$1.91344$\\ \hline
$|z2x2x2z2\rangle$&$1.9875$               &$1.92687$              &$1.99008$              &$1.91503$\\ \hline
\end{tabular}
\caption{Observed maximum and minimum values of $M(e^{-iHt}|S\rangle\langle S|e^{iHt})$ for $L=4$, $(j,h_x,h_z)=(1,-1.05,0.5)$ with $t>100$.
We have chosen as the initial states $|S\rangle$ those state in the form $|I_1n_1\rangle\otimes |I_2n_2\rangle\otimes\cdots |I_Ln_L\rangle$ with $|In\rangle$ the eigenstate of $I$ with eigenvalue $\omega^n$ ($I=x,z$; $n=0,1,\cdots,d$), which we have abbreviated as $|I_1n_1I_2n_2\cdots I_Ln_L\rangle$.}
\label{211223_Manalatetimevaluelist1}
\end{center}
\end{figure}
\begin{figure}
\begin{center}
\begin{tabular}{|c|c|c|c|c|}
\hline
                  &\multicolumn{2}{|c|}{$h_x=1, h_z=0$, open}     &\multicolumn{2}{|c|}{$h_x=1, h_z=0$, periodic}\\ \hline
$|S\rangle$       &$\text{Max}_{t>100}[M]$&$\text{Min}_{t>100}[M]$&$\text{Max}_{t>100}[M]$   &$\text{Min}_{t>100}[M]$\\ \hline
$|x0x0x0x0\rangle$&$1.95063$              &$1.19597$              &$1.97313$                 &$1.02994$                  \\ \hline
$|x0x1x1x0\rangle$&$1.98697$              &$1.85128$              &$1.98538$                 &$1.80405$                  \\ \hline
$|x0x2x2x0\rangle$&$1.98697$              &$1.85128$              &$1.98538$                 &$1.80405$                  \\ \hline
$|x0z0z0x0\rangle$&$1.97238$              &$1.74603$              &$1.97888$                 &$1.65442$                  \\ \hline
$|x0z1z1x0\rangle$&$1.99427$              &$1.73786$              &$1.99098$                 &$1.63055$                  \\ \hline
$|x0z2z2x0\rangle$&$1.97238$              &$1.74603$              &$1.97888$                 &$1.65442$                  \\ \hline
$|x1x0x0x1\rangle$&$1.98562$              &$1.92734$              &$1.98538$                 &$1.80405$                  \\ \hline
$|x1x1x1x1\rangle$&$1.98471$              &$1.68569$              &$1.99113$                 &$1.47506$                  \\ \hline
$|x1x2x2x1\rangle$&$1.98424$              &$1.85524$              &$1.98713$                 &$1.54085$                  \\ \hline
$|x1z0z0x1\rangle$&$1.99302$              &$1.94148$              &$1.98737$                 &$1.90967$                  \\ \hline
$|x1z1z1x1\rangle$&$1.98519$              &$1.88945$              &$1.98748$                 &$1.91064$                  \\ \hline
$|x1z2z2x1\rangle$&$1.98772$              &$1.93549$              &$1.98578$                 &$1.8906$                 \\ \hline
$|x2x0x0x2\rangle$&$1.98562$              &$1.92734$              &$1.98538$                 &$1.80405$                  \\ \hline
$|x2x1x1x2\rangle$&$1.98424$              &$1.85524$              &$1.98713$                 &$1.54085$                  \\ \hline
$|x2x2x2x2\rangle$&$1.98471$              &$1.68569$              &$1.99113$                 &$1.47506$                  \\ \hline
$|x2z0z0x2\rangle$&$1.98772$              &$1.93549$              &$1.98578$                 &$1.8906$                 \\ \hline
$|x2z1z1x2\rangle$&$1.98519$              &$1.88945$              &$1.98748$                 &$1.91064$                  \\ \hline
$|x2z2z2x2\rangle$&$1.99302$              &$1.94148$              &$1.98737$                 &$1.90967$                  \\ \hline
$|z0x0x0z0\rangle$&$1.98424$              &$1.78507$              &$1.97888$                 &$1.65442$                  \\ \hline
$|z0x1x1z0\rangle$&$1.98719$              &$1.91027$              &$1.98737$                 &$1.90967$                  \\ \hline
$|z0x2x2z0\rangle$&$1.98727$              &$1.91742$              &$1.98578$                 &$1.8906$                  \\ \hline
$|z1x0x0z1\rangle$&$1.99068$              &$1.77276$              &$1.99098$                 &$1.63055$                  \\ \hline
$|z1x1x1z1\rangle$&$1.98859$              &$1.90417$              &$1.98748$                 &$1.91064$                  \\ \hline
$|z1x2x2z1\rangle$&$1.98859$              &$1.90417$              &$1.98748$                 &$1.91064$                  \\ \hline
$|z2x0x0z2\rangle$&$1.98424$              &$1.78507$              &$1.97888$                 &$1.65442$                  \\ \hline
$|z2x1x1z2\rangle$&$1.98727$              &$1.91742$              &$1.98578$                 &$1.8906$                  \\ \hline
$|z2x2x2z2\rangle$&$1.98719$              &$1.91027$              &$1.98737$                 &$1.90967$                  \\ \hline
\end{tabular}
\caption{Observed maximum and minimum values of $M(e^{-iHt}|S\rangle\langle S|e^{iHt})$ for $L=4$, $(j,h_x,h_z)=(1,1,0)$ with $t>100$.
}
\label{211223_Manalatetimevaluelist2}
\end{center}
\end{figure}

\newpage

\bibliographystyle{ieeetr}
\bibliography{reference_v2}

\newcommand{\nat}{Nature}\newcommand{\prb}{Phys. Rev. B}\newcommand{\prd}{Phys.
  Rev. D}\newcommand{\pre}{Phys. Rev. E}\newcommand{\prl}{Phys. Rev. Lett.}
\begin{thebibliography}{10}

\bibitem{1999IJTP...38.1113M}
J.~{Maldacena}, ``{The Large-N Limit of Superconformal Field Theories and
  Supergravity},'' {\em International Journal of Theoretical Physics}, vol.~38,
  pp.~1113--1133, Jan 1999.

\bibitem{1998}
S.~Gubser, I.~Klebanov, and A.~Polyakov, ``Gauge theory correlators from
  non-critical string theory,'' {\em Physics Letters B}, vol.~428,
  pp.~105--114, May 1998.

\bibitem{Witten:1998qj}
E.~Witten, ``{Anti-de Sitter space and holography},'' {\em Adv. Theor. Math.
  Phys.}, vol.~2, pp.~253--291, 1998.

\bibitem{2006PhRvL..96r1602R}
S.~{Ryu} and T.~{Takayanagi}, ``{Holographic Derivation of Entanglement Entropy
  from the anti de Sitter Space/Conformal Field Theory Correspondence},'' {\em
  \prl}, vol.~96, p.~181602, May 2006.

\bibitem{Susskind:2014moa}
L.~Susskind, ``{Entanglement is not enough},'' {\em Fortsch. Phys.}, vol.~64,
  pp.~49--71, 2016.

\bibitem{White:2020zoz}
C.~D. White, C.~Cao, and B.~Swingle, ``{Conformal field theories are
  magical},'' {\em Phys. Rev. B}, vol.~103, no.~7, p.~075145, 2021.

\bibitem{Hartman:2013qma}
T.~Hartman and J.~Maldacena, ``{Time Evolution of Entanglement Entropy from
  Black Hole Interiors},'' {\em JHEP}, vol.~05, p.~014, 2013.

\bibitem{2005quant.ph..2070N}
M.~A. {Nielsen}, ``{A geometric approach to quantum circuit lower bounds},''
  {\em arXiv e-prints}, pp.~quant--ph/0502070, Feb. 2005.

\bibitem{2006Sci...311.1133N}
M.~A. {Nielsen}, M.~R. {Dowling}, M.~{Gu}, and A.~C. {Doherty}, ``{Quantum
  Computation as Geometry},'' {\em Science}, vol.~311, pp.~1133--1135, Feb.
  2006.

\bibitem{2007quant.ph..1004D}
M.~R. {Dowling} and M.~A. {Nielsen}, ``{The geometry of quantum computation},''
  {\em arXiv e-prints}, pp.~quant--ph/0701004, Dec. 2006.

\bibitem{Susskind:2014rva}
L.~Susskind, ``{Computational Complexity and Black Hole Horizons},'' {\em
  Fortsch. Phys.}, vol.~64, pp.~24--43, 2016.
\newblock [Addendum: Fortsch.Phys. 64, 44--48 (2016)].

\bibitem{Brown:2015bva}
A.~R. Brown, D.~A. Roberts, L.~Susskind, B.~Swingle, and Y.~Zhao,
  ``{Holographic Complexity Equals Bulk Action?},'' {\em Phys. Rev. Lett.},
  vol.~116, no.~19, p.~191301, 2016.

\bibitem{Brown:2015lvg}
A.~R. Brown, D.~A. Roberts, L.~Susskind, B.~Swingle, and Y.~Zhao,
  ``{Complexity, action, and black holes},'' {\em Phys. Rev. D}, vol.~93,
  no.~8, p.~086006, 2016.

\bibitem{Goto:2018iay}
K.~Goto, H.~Marrochio, R.~C. Myers, L.~Queimada, and B.~Yoshida, ``{Holographic
  Complexity Equals Which Action?},'' {\em JHEP}, vol.~02, p.~160, 2019.

\bibitem{Belin:2021bga}
A.~Belin, R.~C. Myers, S.-M. Ruan, G.~S\'arosi, and A.~J. Speranza,
  ``{Complexity Equals Anything?},'' 11 2021.

\bibitem{Gottesman:1997zz}
D.~Gottesman, ``{Stabilizer codes and quantum error correction},'' 5 1997.

\bibitem{Gottesman:1998hu}
D.~Gottesman, ``{The Heisenberg representation of quantum computers},'' in {\em
  {22nd International Colloquium on Group Theoretical Methods in Physics}},
  pp.~32--43, 7 1998.

\bibitem{Fliss:2020yrd}
J.~R. Fliss, ``{Knots, links, and long-range magic},'' {\em JHEP}, vol.~04,
  p.~090, 2021.

\bibitem{2016PhRvX...6b1043B}
S.~{Bravyi}, G.~{Smith}, and J.~A. {Smolin}, ``{Trading Classical and Quantum
  Computational Resources},'' {\em Physical Review X}, vol.~6, p.~021043, Apr.
  2016.

\bibitem{2016PhRvL.116y0501B}
S.~{Bravyi} and D.~{Gosset}, ``{Improved Classical Simulation of Quantum
  Circuits Dominated by Clifford Gates},'' {\em \prl}, vol.~116, p.~250501,
  June 2016.

\bibitem{2020PhRvB.101q4313C}
B.~{Craps}, M.~{De Clerck}, D.~{Janssens}, V.~{Luyten}, and C.~{Rabideau},
  ``{Lyapunov growth in quantum spin chains},'' {\em \prb}, vol.~101,
  p.~174313, May 2020.

\bibitem{Gottesman:1997qd}
D.~Gottesman, ``{A Theory of fault tolerant quantum computation},'' {\em Phys.
  Rev. A}, vol.~57, p.~127, 1998.

\bibitem{Gottesman:1998se}
D.~Gottesman, ``{Fault tolerant quantum computation with higher dimensional
  systems},'' {\em Chaos Solitons Fractals}, vol.~10, pp.~1749--1758, 1999.

\bibitem{2006JMP....47l2107G}
D.~{Gross}, ``{Hudson's theorem for finite-dimensional quantum systems},'' {\em
  Journal of Mathematical Physics}, vol.~47, pp.~122107--122107, Dec. 2006.

\bibitem{2014NJPh...16a3009V}
V.~{Veitch}, S.~A. {Hamed Mousavian}, D.~{Gottesman}, and J.~{Emerson}, ``{The
  resource theory of stabilizer quantum computation},'' {\em New Journal of
  Physics}, vol.~16, p.~013009, Jan. 2014.

\bibitem{2020arXiv201113937W}
C.~D. {White} and J.~H. {Wilson}, ``{Mana in Haar-random states},'' {\em arXiv
  e-prints}, p.~arXiv:2011.13937, Nov. 2020.

\bibitem{2020PhRvA.102d2409J}
A.~{Jain} and S.~{Prakash}, ``{Qutrit and ququint magic states},'' {\em
  Physical Review A}, vol.~102, p.~042409, Oct. 2020.

\bibitem{2017PhRvL.118i0501H}
M.~{Howard} and E.~{Campbell}, ``{Application of a Resource Theory for Magic
  States to Fault-Tolerant Quantum Computing},'' {\em \prl}, vol.~118,
  p.~090501, Mar. 2017.

\bibitem{2018PhRvA..97f2332A}
M.~{Ahmadi}, H.~B. {Dang}, G.~{Gour}, and B.~C. {Sanders}, ``{Quantification
  and manipulation of magic states},'' {\em Physical Review A}, vol.~97,
  p.~062332, June 2018.

\bibitem{2018arXiv180710296H}
M.~{Heinrich} and D.~{Gross}, ``{Robustness of Magic and Symmetries of the
  Stabiliser Polytope},'' {\em arXiv e-prints}, p.~arXiv:1807.10296, July 2018.

\bibitem{2020NJPh...22h3077S}
S.~{Sarkar}, C.~{Mukhopadhyay}, and A.~{Bayat}, ``{Characterization of an
  operational quantum resource in a critical many-body system},'' {\em New
  Journal of Physics}, vol.~22, p.~083077, Aug. 2020.

\bibitem{RevModPhys.81.865}
R.~Horodecki, P.~Horodecki, M.~Horodecki, and K.~Horodecki, ``Quantum
  entanglement,'' {\em Rev. Mod. Phys.}, vol.~81, pp.~865--942, Jun 2009.

\bibitem{2011arXiv1111.3882B}
F.~G.~S.~L. {Brand{\~a}o}, M.~{Horodecki}, J.~{Oppenheim}, J.~M. {Renes}, and
  R.~W. {Spekkens}, ``{The Resource Theory of Quantum States Out of Thermal
  Equilibrium},'' {\em arXiv e-prints}, p.~arXiv:1111.3882, Nov. 2011.

\bibitem{2013IJMPB..2745019H}
M.~{Horodecki} and J.~{Oppenheim}, ``{(quantumness in the Context Of) Resource
  Theories},'' {\em International Journal of Modern Physics B}, vol.~27,
  p.~1345019, Jan. 2013.

\bibitem{2019RvMP...91b5001C}
E.~{Chitambar} and G.~{Gour}, ``{Quantum resource theories},'' {\em Reviews of
  Modern Physics}, vol.~91, p.~025001, Apr. 2019.

\bibitem{PhysRevA.53.2046}
C.~H. Bennett, H.~J. Bernstein, S.~Popescu, and B.~Schumacher, ``Concentrating
  partial entanglement by local operations,'' {\em Phys. Rev. A}, vol.~53,
  pp.~2046--2052, Apr 1996.

\bibitem{2000JMOp...47..355V}
G.~{Vidal}, ``{Entanglement monotones},'' {\em Journal of Modern Optics},
  vol.~47, pp.~355--376, Feb. 2000.

\bibitem{1984LNP...209....1B}
O.~{Bohigas} and M.-J. {Giannoni}, {\em {Chaotic motion and random matrix
  theories}}, vol.~209, pp.~1--99.
\newblock 1984.

\bibitem{Guhr:1997ve}
T.~Guhr, A.~Muller-Groeling, and H.~A. Weidenmuller, ``{Random matrix theories
  in quantum physics: Common concepts},'' {\em Phys. Rept.}, vol.~299,
  pp.~189--425, 1998.

\bibitem{1977RSPSA.356..375B}
M.~V. {Berry} and M.~{Tabor}, ``{Level Clustering in the Regular Spectrum},''
  {\em Proceedings of the Royal Society of London Series A}, vol.~356,
  pp.~375--394, Sept. 1977.

\bibitem{PhysRevLett.52.1}
O.~Bohigas, M.~J. Giannoni, and C.~Schmit, ``Characterization of chaotic
  quantum spectra and universality of level fluctuation laws,'' {\em Phys. Rev.
  Lett.}, vol.~52, pp.~1--4, Jan 1984.

\bibitem{1969JETP...28.1200L}
A.~I. {Larkin} and Y.~N. {Ovchinnikov}, ``{Quasiclassical Method in the Theory
  of Superconductivity},'' {\em Soviet Journal of Experimental and Theoretical
  Physics}, vol.~28, p.~1200, June 1969.

\bibitem{Nosaka:2018iat}
T.~Nosaka, D.~Rosa, and J.~Yoon, ``{The Thouless time for mass-deformed SYK},''
  {\em JHEP}, vol.~09, p.~041, 2018.

\bibitem{Kudler-Flam:2019kxq}
J.~Kudler-Flam, L.~Nie, and S.~Ryu, ``{Conformal field theory and the web of
  quantum chaos diagnostics},'' {\em JHEP}, vol.~01, p.~175, 2020.

\bibitem{2011PhRvL.106e0405B}
M.~C. {Ba{\~n}uls}, J.~I. {Cirac}, and M.~B. {Hastings}, ``{Strong and Weak
  Thermalization of Infinite Nonintegrable Quantum Systems},'' {\em \prl},
  vol.~106, p.~050405, Feb 2011.

\bibitem{Page:1993df}
D.~N. Page, ``{Average entropy of a subsystem},'' {\em Phys. Rev. Lett.},
  vol.~71, pp.~1291--1294, 1993.

\bibitem{Sen:1996ph}
S.~Sen, ``{Average entropy of a subsystem},'' {\em Phys. Rev. Lett.}, vol.~77,
  pp.~1--3, 1996.

\bibitem{Goto:2021gve}
K.~Goto, A.~Mollabashi, M.~Nozaki, K.~Tamaoka, and M.~T. Tan, ``{Information
  Scrambling Versus Quantum Revival Through the Lens of Operator
  Entanglement},'' 12 2021.

\bibitem{2020arXiv201214609M}
E.~{Mascot}, M.~{Nozaki}, and M.~{Tezuka}, ``{Local Operator Entanglement in
  Spin Chains},'' {\em arXiv e-prints}, p.~arXiv:2012.14609, Dec. 2020.

\end{thebibliography}

\end{document}